\documentclass[review]{elsarticle}
\pdfoutput=1
\usepackage{lineno}
\usepackage{graphicx,amssymb,arydshln,subeqnarray,subfigure}
\usepackage{url}
\usepackage{amsbsy,amsmath,amsfonts,cases,bm,caption}
\journal{Signal Processing}

\begin{document}

\begin{frontmatter}

\title{TDOA-based localization with NLOS mitigation via robust model transformation and\\neurodynamic optimization}

\author[mymainaddress]{Wenxin~Xiong\corref{mycorrespondingauthor}}
\cortext[mycorrespondingauthor]{Corresponding author}
\ead{w.x.xiong@outlook.com}

\author[mymainaddress]{Christian~Schindelhauer}
\ead{schindel@informatik.uni-freiburg.de}

\author[mysecondaryaddress]{Hing~Cheung~So}
\ead{hcso@ee.cityu.edu.hk}

\author[mymainaddress]{Joan~Bordoy}
\ead{bordoy@informatik.uni-freiburg.de}

\author[mymainaddress]{Andrea~Gabbrielli}
\ead{andre\_gabe@hotmail.it}

\author[mythirdaddress]{Junli~Liang}
\ead{liangjunli@nwpu.edu.cn}

\address[mymainaddress]{Department of Computer Science, University of Freiburg, Freiburg 79110, Germany}
\address[mysecondaryaddress]{Department of Electrical Engineering, City University of Hong Kong, Hong Kong, China}
\address[mythirdaddress]{School of Electronics and Information, Northwestern Polytechnical University, Xi'an 710072, China}

\begin{abstract}
This paper revisits the problem of locating a signal-emitting source from time-difference-of-arrival (TDOA) measurements under non-line-of-sight (NLOS) propagation. Many currently fashionable methods for NLOS mitigation in TDOA-based localization tend to solve their optimization problems by means of convex relaxation and, thus, are computationally inefficient. Besides, previous studies show that manipulating directly on the TDOA metric usually gives rise to intricate estimators. Aiming at bypassing these challenges, we turn to retrieve the underlying time-of-arrival framework by treating the unknown source onset time as an optimization variable and imposing certain inequality constraints on it, mitigate the NLOS errors through the $\ell_1$-norm robustification, and finally apply a hardware realizable neurodynamic model based on the redefined augmented Lagrangian and projection theorem to solve the resultant nonconvex optimization problem with inequality constraints. It is validated through extensive simulations that the proposed scheme can strike a nice balance between localization accuracy, computational complexity, and prior knowledge requirement.
\end{abstract}


\end{frontmatter}


\section{Introduction}
Source localization using measurements from spatially separated passive sensors has turned into a go-to scheme in many location-based services including target tracking \cite{FHoeflinger,JBordoy}, human-computer interaction \cite{VGReju}, and Internet of Things \cite{SLi}. Among plentiful measurement models, the time-of-arrival (TOA) and time-difference-of-arrival (TDOA), especially the latter that eliminates the need for synchronization between the source and sensors \cite{YHuang,NOno1,NOno2,LLin}, is perhaps the most widely used owing to its high accuracies. For an insight into the rationale of single source localization, the uninitiated readers are referred to \cite{So,Guvenc1} and the references therein. 

One of the key issues in source localization is the so-called non-line-of-sight (NLOS) propagation, which commonly arises in real environments (e.g., urban canyons and indoor sites), and can adversely degrade the positioning performance if left untreated \cite{Guvenc1,STomic1,STomic2,GWang,HChen,WXiong,AProrok,GWang2,WWang,GWang3,ZSu,WXiong2}. Over the past decade, a vast variety of advanced NLOS mitigation methods have been developed for TOA-based localization: the worst-case least squares (LS) \cite{STomic1}, joint estimation of the source location and a balancing parameter \cite{STomic2,GWang,HChen}, and robust multidimensional similarity analysis \cite{WXiong}, to name a few. These approaches are practically more favorable than the straightforward maximum likelihood (ML) technique \cite{AProrok}, as their implementations rely on neither the path status nor the specified error distribution, but merely a few assumptions regarding the measurement noise and/or NLOS errors. Different from what one might expect, extension of the aforementioned TOA-based schemes to the TDOA case is not at all a trivial task. This is mainly because the possible NLOS error in a TDOA measurement is essentially the difference of those occurred in two related TOA measurements, and hence may not necessarily be a positive outlier anymore. To settle this matter, the authors of \cite{GWang2} follow again the worst-case rule, but this time the upper bound is imposed on the magnitude of NLOS errors. As a modification to \cite{GWang2} which treats each measurement equally, additional path status information is utilized in \cite{WWang} for placing less reliance on the error-prone measurements. More recently, the authors of \cite{GWang3} point out that the formulations in \cite{GWang2} and \cite{WWang} may not perform well due to the loose upper bound and inexact triangle inequality, whereupon they put forward several refinements to alleviate the impacts. Despite considerable resistance of the worst-case criterion to NLOS errors, solving the resultant robust LS problems in \cite{GWang2,WWang,GWang3}, however, involves the use of convex optimization such as second-order cone programming (SOCP) and semidefinite programming (SDP), which will bring in heavy computational burdens. On the other hand, whereas the TDOA model with a structure more complex than the TOA counterpart can impede the formulation derivation \cite{TLe}, the idea of model transformation is suggested in \cite{ZSu,EXu}. Such a tactic is well-motivated to the extent that the metrics of TOA and TDOA differ by only one degree of freedom, i.e., the time at which the signal departs from the source. Moreover, the selection of a proper reference sensor is no longer a prerequisite after the model transformation. Nevertheless, the constrained LS estimator with NLOS mitigation in \cite{ZSu} still ends up with solving a complicated SDP problem.

Conventional numerical methods for optimization are often realized and run on digital computers. Consequently, the computing time can grow dramatically with the increase of problem size, implying less effectiveness in time-varying scenarios. To overcome this drawback, employing physically implementable recurrent neural networks by which distributed, parallel, and real-time computation is enabled has become a promising alternative for tackling various classes of mathematical programming problems \cite{SZhang,DTank,MPKennedy,ANazemi,XHu,ZShi2,HChe}. The mechanism is to build a dynamical system that will ultimately settle down to an equilibrium point, at which the optimal solution to the problem is obtained from the outputs, given suitable inputs as the initial point. In particular, the Lagrange programming neural networks (LPNN) \cite{SZhang} developed based on the gradient model \cite{MPKennedy} and Lagrange multiplier theory has provided a general framework for coping with the nonlinear constrained optimization problems. With the use of an augmented Lagrangian function, the LPNN model can further be empowered to handle nonconvex optimization, and recent studies have successfully utilized the augmented LPNN to solve a mass of source localization problems \cite{HWang,ZHan,JLiang,ZFHan,CJia}. However, the standard LPNN framework is unfriendly towards the presence of inequality constraints, since it requires introducing slack variables to convert them into the equality ones as a preprocessing step. This is apparently not a fine option if a large number of inequality constraints are involved, in view of the fact that promptness and real-time responses are the main purposes of applying the recurrent neural networks. Unfortunately, for the sake of binding the additional nuisance variable, there do exist many inequality constraints in the model transformation approaches \cite{ZSu}, which indicates that a more efficient means of neurodynamic optimization is still a yearning in our application.

In this paper, we formulate TDOA-based source localization in NLOS environments as a nonconvex constrained optimization problem by robust model transformation, and then devise an effective and efficient neurodynamic solution to it. To start with, the least absolute deviation (LAD) (also known as (a.k.a.) the $\ell_1$-norm) criterion is adopted to achieve robustness against the bias-like NLOS error in the reconstructed TOA measurement model. For the higher-order properties in the design of dynamical system, certain smoothed approximations are made to the LAD objective function to yield a twice differentiable surrogate. Unlike most of the neurodynamic source localization approaches adapting their formulations to the standard LPNN setting (e.g., by either discarding the inequality constraints \cite{HWang,JLiang} or transforming them into the equalities \cite{ZHan,ZFHan,CJia}), we follow \cite{XHu} to redefine the augmented Lagrangian and establish a different projection-type neural network (PNN) model which can directly take the inequality constraints into account. It is worth noting that although the LPNN and PNN share the same terminology ``neural network'' with the booming deep neural networks in machine learning, they refer to totally distinct approaches and should not be mixed up with each other. The presented scheme obviates the need for acquiring any information (e.g., an upper bound \cite{GWang2,WWang,GWang3}) concerning NLOS errors or tuning the hyperparameters \cite{ZSu} beforehand, thereby resulting in a lower prior knowledge demand compared to the methods in \cite{GWang2,WWang,GWang3,ZSu}. It should be noted that though bearing some resemblance to \cite{ZSu} which also remodels the problem into a TOA framework, our work should be distinguished from it, as neither the $\ell_2$-space-based objective function nor the time-consuming SDP is counted on any longer. In addition, our neurodynamic solution is shown to be computationally more efficient even when it is executed on the general purpose digital computers.

The remainder of the paper is organized as follows. Section \ref{PF} states the localization problem and introduces the robust model transformation formulation. Section \ref{NN} reviews the classical LPNN framework and defines the neural dynamics of the presented PNN, whose stability and convergence properties are then briefly discussed in Section \ref{CSCA}. To ensure a fair comparison between the proposed neurodynamic method and the state-of-the-art convex optimization counterparts in terms of computational expense, its algorithmic complexity when implementing in a numerical fashion is also analyzed in Section \ref{CSCA}. Section \ref{SR} evaluates the performance of our approach through computer simulations. Finally, conclusions are drawn in Section \ref{CC}.

\section{Problem formulation}
\label{PF}
\begin{figure}[!t]
	\centering
	\includegraphics[width=3.5in]{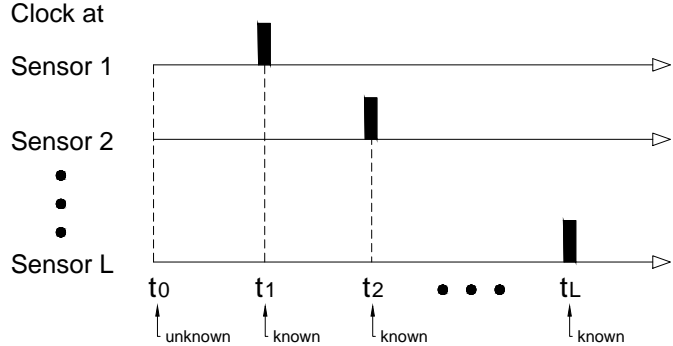}
	\caption{Signal timestamp diagram of TDOA-based localization system.} 
	\label{FIG1}
\end{figure}
Consider a TDOA-based localization system in $k$-dimensional space ($k = 2$ or $3$) with $L \geq k+1$ sensors and a single source. The known sensor positions and unknown source location are denoted by $\bm{x}_i \in \mathbb{R}^k$ (for $i = 1,2,...,L$) and $\bm{x} \in \mathbb{R}^k$, respectively. As demonstrated in Fig. \ref{FIG1}, the local clocks of the sensors are well synchronized such that the received signal timestamp $t_i$ (for $i = 1,2,...,L$) can be collected from the $i$th sensor, whereas the time at which the signal is emitted from the source, $t_0$, is unknown because there is no synchronization between the source and sensors. Without loss of generality, the first sensor is designated as the reference and the TDOA measurements are modeled as
\begin{align}{\label{TDOA}}
t_{i,1} = \frac{1}{c} ({\| \bm{x} - \bm{x}_i \|}_2 - {\| \bm{x} - \bm{x}_1 \|}_2 + n_{i,1} + b_{i,1}) = t_i - t_1,~~i = 2,3,...,L,
\end{align}
where $c$ denotes the signal propagation speed, ${\| \cdot \|}_2$ represents the $\ell_2$-norm of a vector, $n_{i,1} = n_i - n_1$ and $b_{i,1} = q_i - q_1$ (both for $i=2,3,...,L$) are the measurement noise and possible NLOS error in the corresponding range difference measurement, respectively, $n_i$ (for $i = 1,2,...,L$) is assumed to be zero-mean Gaussian noise with variance $\sigma_i^2$, and $q_i$ (for $i = 1,2,...,L$) equals either $0$ or a positive bias error $e_i$, contingent on whether the path between the $i$th sensor and source is line-of-sight (LOS) or NLOS. Before proceeding with the formulation derivation, we decompose the TDOA measurements in (\ref{TDOA}) into the related TOA components
\begin{equation}
t_i - t_0 = \frac{1}{c}({\| \bm{x} - \bm{x}_i \|}_2 + n_{i} + q_{i}),~~i = 1,2,...,L\label{TOA}
\end{equation}
by making as if the synchronization between the source and sensors is established, namely, including $t_0$ as a variable of interest.

Exhibiting less sensitivity to outliers than the conventional $\ell_2$-norm criterion, the $\ell_1$-norm has been widely utilized in robust signal processing, with low-rank matrix completion under impulsive noise circumstances \cite{AEriksson}, robust principal component analysis \cite{ECandes}, and sensor network localization under Laplacian noise assumption \cite{POE} being a few representative applications of it. Borrowing the similar idea, we employ the LAD cost function as the objective of minimization to mitigate the positive bias errors in (\ref{TOA}):
\begin{equation*}
\min_{t_0, \bm{x}}~\sum_{i=1}^{L} \left| (t_i - t_0) c - {\| \bm{x} - \bm{x}_i \|}_2 \right|.
\end{equation*}
To bind the nuisance variable $t_0$, the temporal constraints\footnote{The temporal constraints are premised on $t_i > 0$ so as to be meaningful.}
\begin{equation}
0 \leq t_0 \leq t_i,~~i = 1,2,...,L,\label{TEMC}
\end{equation}
geometrical constraints by the triangle inequalities \cite{OJean}
\begin{equation}
(t_i - t_0)c + (t_j - t_0)c \geq {\| \bm{x}_i - \bm{x}_j \|}_2,~~i \not= j,~i,j = 1,2,...,L,\label{TRIC}
\end{equation}
and general consensus that $e_i$ is much greater than $|n_i|$ are thereupon incorporated into the formulation, yielding:
\begin{subequations}{\label{CLAD}}
	\begin{align}
	\min_{t_0, \bm{x}, \bm{d}} &~\sum_{i=1}^{L} \left| (t_i - t_0) c - d_i \right|\nonumber\\
	\textup{s.t.}~~&d_i^2 = {\| \bm{x} - \bm{x}_i \|}_2^2,~~i = 1,2,...,L,\\
	&d_i \geq 0,~~i = 1,2,...,L,\\
	&\text{(\ref{TEMC})},~\text{(\ref{TRIC})},\nonumber\\
	&(t_i - t_0) c \geq d_i,~~i = 1,2,...,L,
	\end{align}
\end{subequations}
where $\bm{d} = \left[ d_1,d_2,...,d_L \right]^T \in \mathbb{R}^L$ is a vector containing the auxiliary variables for source-sensor distances, and the constraint $d_i = {\| \bm{x} - \bm{x}_i \|}_2$ (for $i = 1, 2, ..., L$) is replaced by (\ref{CLAD}a) and (\ref{CLAD}b) in the quadratic form to avoid ill-posing \cite{ZHan}. Falling into the category of nonlinear and nonconvex constrained optimization problems, (\ref{CLAD}) is appropriately tackled in the next section by constructing a dynamical system whose equilibrium state is reached at a Karush-Kuhn-Tucker (KKT) point of the underlying problem.

\section{Preliminaries and proposed neurodynamic method}
\label{NN}
Assume that we have a nonlinear programming problem with equality constraints:
\begin{align}{\label{ENP}}
\min_{\bm{z}}~f(\bm{z}),\quad\textup{s.t.}~~\bm{h}(\bm{z}) = \bm{0}_M,
\end{align}
where $\bm{z}\in \mathbb{R}^N$, $f:\mathbb{R}^N \rightarrow \mathbb{R}$, $\bm{h}(\bm{z}) = \left[ h_1 (\bm{z}),h_2 (\bm{z}),...,h_M (\bm{z}) \right]^T \in \mathbb{R}^M$ is an $M$-dimensional vector-valued function of $N$ variables with $M \leq N$, the functions $f(\bm{z})$ and $h_i (\bm{z})$ (for $i = 1,2,...,M$) are supposed to be twice differentiable, and $\bm{0}_M \in \mathbb{R}^M$ denotes an all-zero vector of length $M$. In a nutshell, the widely used LPNN approach \cite{SZhang} deals with (\ref{ENP}) by invoking the Lagrange multiplier theory and designing a neurodynamic model whose time-domain transient behavior is defined as
\begin{subequations}{\label{dynamicsENP}}
	\begin{align}
	\frac{d\bm{z}}{dt} &= -\bm{\nabla}_{\bm{z}} \mathcal{L}_{\star}(\bm{z},\bm{\lambda}),\\
	\frac{d{\bm{\lambda}}}{dt} &= \bm{\nabla}_{\bm{\lambda}} \mathcal{L}_{\star}(\bm{z},\bm{\lambda}),
	\end{align}
\end{subequations}
where $\bm{\nabla}_{\bm{z}} (\cdot) \in \mathbb{R}^N$ denotes the gradient of a function at $\bm{z}$, $\bm{\lambda} \in \mathbb{R}^M$ is a vector containing the Lagrange multipliers for the constraints in (\ref{ENP}), $\bm{z}$ and $\bm{\lambda}$ are assigned physical meanings as the activities of the variable and Lagrangian neurons, respectively, and $\mathcal{L}_{\star}(\bm{z},\bm{\lambda})$ can be either the Lagrangian or augmented Lagrangian of (\ref{ENP}), differing in the stability of the built system under nonconvexity. In the dynamic process of the LPNN, (\ref{dynamicsENP}a) ensures that the value of $\mathcal{L}_{\star}(\bm{z},\bm{\lambda})$ decreases over time, whereas (\ref{dynamicsENP}b) plays a role in leading the solution into the feasible region. After performing appropriate initialization of the variable and Lagrangian neurons, the network governed by (\ref{dynamicsENP}) is expected to approach an equilibrium point satisfying the first-order necessary conditions of optimality (a.k.a. the KKT conditions). 

It is obvious that (\ref{CLAD}) does not conform to the paradigm shown in (\ref{ENP}) owing to the existence of numerous inequality constraints. Instead of introducing slack variables \cite{ZHan,ZFHan} to fit in with (\ref{ENP}), in the following we seek for a simpler way to directly handle the general constrained optimization problem (GCOP)
\begin{align}{\label{GCOP}}
\min_{\bm{z}}~f(\bm{z}),\quad\textup{s.t.}~~\bm{g}(\bm{z}) \leqq \bm{0}_K,~~\bm{h}(\bm{z}) = \bm{0}_M,
\end{align}
where the definitions pertaining to $\bm{z}$, $\bm{\lambda}$, $f$, and $\bm{h}$ remain the same as those in (\ref{ENP}) except that $M \leq N$ is no longer requested, the $K$-dimensional vector-valued function $\bm{g}(\bm{z}) = \left[ g_1 (\bm{z}),g_2 (\bm{z}),...,g_K (\bm{z}) \right]^T \in \mathbb{R}^K$ is assumed to be twice differentiable, and the vector inequality $\bm{a} \leqq \bm{b}$ means each component of $\bm{a}$ is less than or equal to each corresponding component of $\bm{b}$.

The Lagrangian of (\ref{GCOP}) is $\mathcal{L}(\bm{z},\bm{\nu}) = f(\bm{z}) + \bm{\mu}^T \bm{g}(\bm{z}) + \bm{\lambda}^T \bm{h}(\bm{z})$. Here, we have $\bm{\nu} = \left[ \bm{\mu}^T, \bm{\lambda}^T \right]^T \in \mathbb{R}^{K+M}$, where $\bm{\mu} = \left[ \mu_1,\mu_2,...,\mu_K \right]^T \in \mathbb{R}^K$ and $\bm{\lambda} = \left[ \lambda_1,\lambda_2,...,\lambda_M \right]^T \in \mathbb{R}^M$ are the vectors containing Lagrange multipliers for the inequality and equality constraints in (\ref{GCOP}), respectively. The KKT conditions \cite{JNocedal} for (\ref{GCOP}) that a pair $(\bm{z}^{*},\bm{\nu}^{*})$ satisfies\footnote{In this paper, we stipulate that the asterisk in the superscript of a vector is by default applied to each component of the vector.}, namely, the first-order necessary conditions for $\bm{z}^{*}$ to be a local minimizer of (\ref{GCOP}), are
\begin{subequations}{\label{KKT}}
	\begin{numcases}{}
	\bm{\nabla}_{\bm{z}} \mathcal{L}(\bm{z}^{*},\bm{\nu}^{*}) = \bm{0}_N,\\
	g_i (\bm{z}^{*}) \leq 0, \mu_i^{*} \geq 0, \mu_i^{*} g_i (\bm{z}^{*}) = 0,~~i=1,2,...,K,\\
	\bm{h}(\bm{z}^{*}) = \bm{0}_M.
	\end{numcases}
\end{subequations}
Analogous to the strategy taken by \cite{XHu}, we point out that the KKT conditions in (\ref{KKT}) actually share the same solution set with
\begin{subequations}{\label{KKT2}}
	\begin{numcases}{}
	\bm{\nabla}_{\bm{z}} \mathcal{L}_{\rho} (\bm{z}^{*},\bm{\nu}^{*}) = \bm{0}_N,\\
	\left[ \mu_{i}^{*} + \alpha g_i (\bm{z}^{*}) \right]^{+} = \mu_{i}^{*},~~i=1,2,...,K,\\
	\bm{h}(\bm{z}^{*}) = \bm{0}_M,
	\end{numcases}
\end{subequations}
where $\mathcal{L}_{\rho}(\bm{z},\bm{\nu}) = \mathcal{L}(\bm{z},\bm{\nu}) + \frac{\rho}{2} \Big\{ \sum_{i = 1}^{K} \left[ \mu_{i} g_i (\bm{z}) \right]^2 + \sum_{i = 1}^{M} \left[ \lambda_{i} h_i (\bm{z}) \right]^2 \Big\}$ is a redefined augmented Lagrangian of (\ref{GCOP}), the scale factor $\alpha > 0$ indicates the convergence rate of the neural network and we let $\alpha = 1$ in this paper for simplicity, $\rho > 0$ is the augmented Lagrangian parameter, and the operator
\begin{equation}{\label{PO}}
\left[ \cdot \right]^{+} = \max(\cdot,0)
\end{equation}
is introduced to re-express the primal feasibility, dual feasibility, and complementarity conditions for the inequality constraints in a projection form. The equivalence between the solution sets of (\ref{KKT}) and (\ref{KKT2}) is illustrated in the proposition below.

\textbf{Proposition 1.} Denote the solution sets of equations in (\ref{KKT}) and (\ref{KKT2}) by $\Omega_1$ and $\Omega_2$, respectively, then $\Omega_1 = \Omega_2$.

\textbf{Proof.} We begin with proving that (\ref{KKT}b) is true if and only if (\ref{KKT2}b) is true.

\textit{Sufficiency:}

The conditions in (\ref{KKT}b) are partitioned into two cases as: (i) $g_i (\bm{z}^{*}) = 0, \mu_i^{*} \geq 0$, and (ii) $g_i (\bm{z}^{*}) < 0, \mu_i^{*} = 0$. The equalities in (\ref{KKT2}b) can be trivially deduced in both two cases, thus the sufficiency holds.

\textit{Necessity:}

\textit{Case 1:} $\mu_{i}^{*} + \alpha g_i (\bm{z}^{*}) \geq 0$.

It follows from (\ref{KKT2}b) and (\ref{PO}) that $\left[ \mu_{i}^{*} + \alpha g_i (\bm{z}^{*}) \right]^{+} = \mu_{i}^{*} + \alpha g_i (\bm{z}^{*}) = \mu_{i}^{*}$, which subsequently implies $g_i (\bm{z}^{*}) = 0$ and $\mu_{i}^{*} \geq 0$.

\textit{Case 2:} $\mu_{i}^{*} + \alpha g_i (\bm{z}^{*}) < 0$.

Likewise, we arrive at $\left[ \mu_{i}^{*} + \alpha g_i (\bm{z}^{*}) \right]^{+} = 0 = \mu_{i}^{*}$ and $g_i (\bm{z}^{*}) < 0$.

It is evident that the conditions in (\ref{KKT}b) are formed by merging the two cases together. Therefore, the necessity is satisfied.

In this way, we now only need to prove that (\ref{KKT}a) and (\ref{KKT2}a) are equivalent to each other under the conditions in (\ref{KKT}b) and (\ref{KKT}c). The gradient of $\mathcal{L}_{\rho}(\bm{z},\bm{\nu})$ at $\bm{z}$ is calculated as
\begin{align}{\label{gradLrho}}
\bm{\nabla}_{\bm{z}} \mathcal{L}_{\rho} (\bm{z},\bm{\nu}) = \bm{\nabla}_{\bm{z}} \mathcal{L} (\bm{z},\bm{\nu}) + \rho \Bigg[ \sum_{i = 1}^{K} \mu_{i}^2 g_i (\bm{z}) \bm{\nabla}_{\bm{z}} g_i (\bm{z}) + \sum_{i = 1}^{M} \lambda_{i}^2 h_i (\bm{z}) \bm{\nabla}_{\bm{z}} h_i (\bm{z}) \Bigg].
\end{align}
Substituting the conditions in (\ref{KKT}b) and (\ref{KKT}c) into (\ref{gradLrho}) at $(\bm{z}^{*},\bm{\nu}^{*})$ produces $\bm{\nabla}_{\bm{z}} \mathcal{L}_{\rho} (\bm{z}^{*},\bm{\nu}^{*}) = \bm{\nabla}_{\bm{z}} \mathcal{L} (\bm{z}^{*},\bm{\nu}^{*})$, which verifies the equivalence between (\ref{KKT}a) and (\ref{KKT2}a). The proof is complete.
\begin{figure}[!t]
	\centering
	\includegraphics[width=3.5in]{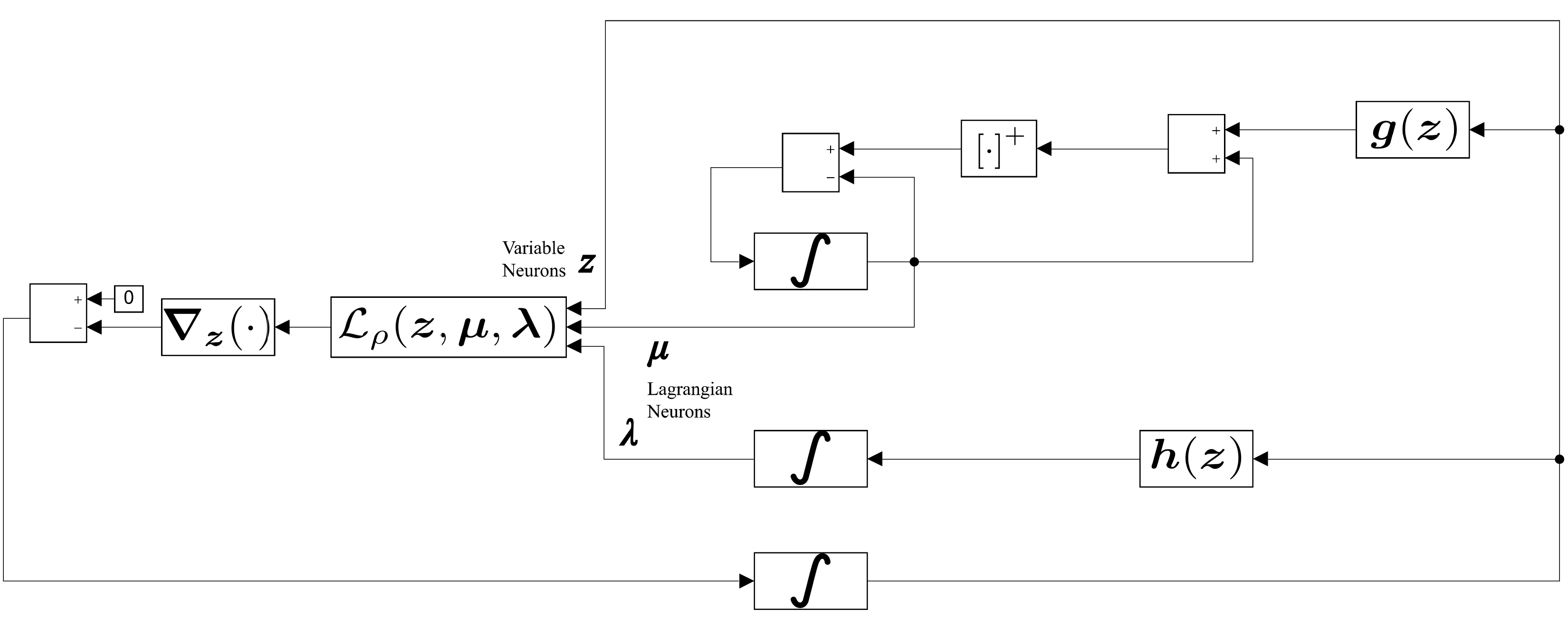}
	\caption{Sketch for neural network defined by (\ref{dynamicsGCOP}).} 
	\label{xiong2}
\end{figure}

Based on (\ref{KKT2}), a KKT point of the GCOP (\ref{GCOP}) is to be searched by employing a three-layer PNN, with its dynamical equations being given by
\begin{subequations}{\label{dynamicsGCOP}}
	\begin{align}
	\frac{d\bm{z}}{dt} &= -\bm{\nabla}_{\bm{z}} \mathcal{L}_{\rho} (\bm{z},\bm{\nu}),\\
	\frac{d{\mu_{i}}}{dt} &= -\mu_{i} + \left[ \mu_{i} + g_i (\bm{z}) \right]^{+},~~i=1,2,...,K,\\
	\frac{d{\bm{\lambda}}}{dt} &= \bm{h} (\bm{z}).
	\end{align}
\end{subequations}
A simplified block diagram of how such a neural network can be implemented on hardware is sketched in Fig. \ref{xiong2}. What may be noteworthy is that (\ref{dynamicsGCOP}) can be viewed as either a projection-type extension of the standard LPNN \cite{SZhang}, a GCOP-treatable augmentation of the neurodynamic model in \cite{XHu}, or a simplification leaving out the bound constraints of that in \cite{HChe}. On this account, several existing analyses in the literature will be referenced for the property discussion on (\ref{dynamicsGCOP}) in the related sections.

\begin{figure}[!t]
	\centering
	\includegraphics[width=3.5in]{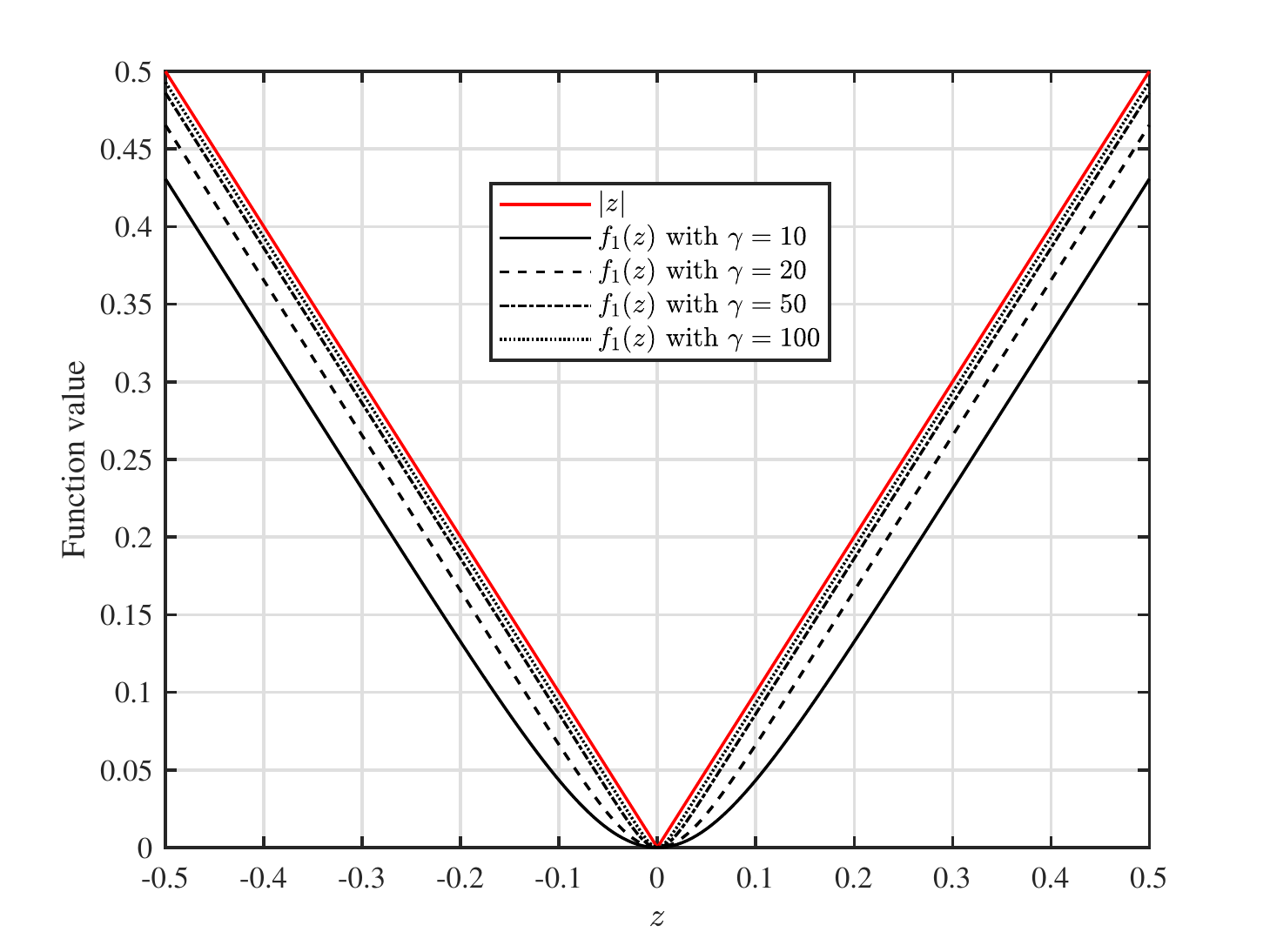}
	\caption{Comparison of the functions $|z|$ and $f(z)$.} 
	\label{xiong3}
\end{figure}

In what follows, the neurodynamic system described by (\ref{dynamicsGCOP}) is exploited for working out the solution to (\ref{CLAD}). To meet the higher-order (more precisely, twice in our scenario) differentiability condition for the neural network implementation \cite{SZhang,HChe}, the absolute value function in (\ref{CLAD}) is replaced by the following smoothed robust loss function with arbitrary-order derivatives\footnote{Note that the celebrated Huber loss function which is a trade-off between the $\ell_1$- and $\ell_2$-norm \cite{RZhang} also suffers from the differentiability issues, i.e., it is only first-order differentiable \cite{KFount}.} \cite{LZhao}:
\begin{equation*}
f_1 (z) = \frac{\ln \left( \left( e^{\gamma z} + e^{-\gamma z} \right) / 2 \right)}{\gamma},
\end{equation*}
where $\gamma > 0$ is a predefined parameter and $\log(\cdot)$ denotes the logarithm operation with base $e$. For illustrative purpose, the comparison between the absolute value function $|z|$ and $f_1 (z)$ is provided in Fig. \ref{xiong3}, from which it is clearly seen that acceptable approximation can be achieved if a sufficiently large $\gamma$ is chosen.
Accordingly, the problem (\ref{CLAD}) is approximated by
\begin{align*}
\min_{t_0, \bm{x}, \bm{d}} 	~\sum_{i=1}^{L} f_1 ((t_i - t_0) c - d_i), \quad \textup{s.t.}~~\text{(\ref{TEMC})},~\text{(\ref{TRIC})},~\text{(\ref{CLAD}a)--(\ref{CLAD}c)},
\end{align*}
which can then be cast into the standard GCOP form shown in (\ref{GCOP}) by letting 
\begin{subequations}
	\begin{align*}
	&\bm{z} = \left[ t_0, \bm{x}^T, \bm{d}^T \right]^T \in \mathbb{R}^{L+k+1},\\
	&N = L+k+1,\\
	&K = \frac{L^2 + 5L + 2}{2},\\
	&M = L,\\
	&f(\bm{z}) = \sum_{i=1}^{L} f_1 ((t_i - t_0) c - d_i),\\
	&g_1 (\bm{z}) = - t_0,\\
	&g_{i+1} (\bm{z}) = t_0 - t_i,~~i = 1,2,...,L,
	\end{align*}
\end{subequations}
\begin{subequations}
	\begin{align*}
	&g_{i+L+1} (\bm{z}) = -d_i,~~i = 1,2,...,L,\\
	&g_{i+2L+1} (\bm{z}) = d_i - (t_i - t_0)c,~~i = 1,2,...,L,\\
	&\left[ \bm{g}(\bm{z}) \right]_{3L+2:K} =\left[g_{3L+2} (\bm{z}),..., g_{\frac{(2L-i)(i-1)}{2}+j-i+3L+1} (\bm{z}),...,g_{K} (\bm{z})\right]^T\\
	&~~=\left[ g_{1,2} (\bm{z}),...,g_{1,L} (\bm{z}),g_{2,3} (\bm{z}),...,g_{L-1,L} (\bm{z}) \right]^T \in \mathbb{R}^{\frac{L(L-1)}{2}}\\
	&h_i (\bm{z}) = d_i^2 - {\| \bm{x} - \bm{x}_i \|}_2^2,~~i = 1,2,...,L,
	\end{align*}
\end{subequations}
where
\begin{align*}
g_{i,j} (\bm{z}) = (2t_0 - t_i - t_j)c + {\| \bm{x}_i - \bm{x}_j \|}_2,~~i = 1,2,...,L-1,~j = i+1, i+2,...,L.
\end{align*}

While the dynamical equations are readily constructed pursuant to the rules in (\ref{dynamicsGCOP}), a more detailed description of the most crucial step (\ref{dynamicsGCOP}a) is presented as follows:
\begin{subequations}
	\begin{align*}
	\frac{d\bm{z}}{dt} &= \left[ \frac{d t_0}{dt}, \left(\frac{d\bm{x}}{dt}\right)^T, \left(\frac{d\bm{d}}{dt}\right)^T \right]^T = -\bm{\nabla}_{\bm{z}} \mathcal{L}_{\rho} (\bm{z},\bm{\nu}) = -\frac{\partial \mathcal{L}_{\rho} (\bm{z},\bm{\nu})}{\partial \bm{z}}\\
	&= -\left[ \frac{\partial \mathcal{L}_{\rho} (\bm{z},\bm{\nu})}{\partial t_0}, \left(\frac{\partial \mathcal{L}_{\rho} (\bm{z},\bm{\nu})}{\partial \bm{x}}\right)^T, \left(\frac{\partial \mathcal{L}_{\rho} (\bm{z},\bm{\nu})}{\partial \bm{d}}\right)^T \right]^T,
	\end{align*}
\end{subequations}
where
\begin{subequations}
	\begin{align*}
	&\frac{\partial \mathcal{L}_{\rho} (\bm{z},\bm{\nu})}{\partial t_0} = c \sum_{i=1}^{L} \frac{e^{2 \gamma \left[ d_i + (t_0 - t_i)c \right]} - 1}{e^{2 \gamma \left[ d_i + (t_0 - t_i)c \right]} + 1} - \mu_1 + \sum_{i=1}^{L} \mu_{i+1} + c \sum_{i=1}^{L} \mu_{i+2L+1}\\
	&~~+ 2c \sum_{i=1}^{L-1} \sum_{j=i+1}^{L} \mu_{\frac{(2L-i)(i-1)}{2}+j-i+3L+1} + \rho \Bigg\{ \mu_1^2 t_0 + \sum_{i=1}^{L} \mu_{i+1}^2 (t_0 - t_i)\\
	&~~+ c \sum_{i=1}^{L} \mu_{i+2L+1}^2 \left[ d_i - (t_i - t_0) c \right] + 2c \sum_{i=1}^{L-1} \sum_{j=i+1}^{L} \mu_{\frac{(2L-i)(i-1)}{2}+j-i+3L+1}^2\\
	&~~\left[ (2t_0 - t_i - t_j)c + {\| \bm{x}_i - \bm{x}_j \|}_2 \right] \Bigg\},\\
	&\frac{\partial \mathcal{L}_{\rho} (\bm{z},\bm{\nu})}{\partial \bm{x}} = 2 \sum_{i=1}^{L} \left[ \lambda_{i} + \rho \lambda_{i}^2 \left(d_i^2 - {\| \bm{x} - \bm{x}_i \|}_2^2 \right) \right] \left( \bm{x}_i - \bm{x} \right),\\
	&\frac{\partial \mathcal{L}_{\rho} (\bm{z},\bm{\nu})}{\partial \bm{d}} = \left[ \frac{\partial \mathcal{L}_{\rho} (\bm{z},\bm{\nu})}{\partial d_1}, \frac{\partial \mathcal{L}_{\rho} (\bm{z},\bm{\nu})}{\partial d_2},..., \frac{\partial \mathcal{L}_{\rho} (\bm{z},\bm{\nu})}{\partial d_L} \right]^T,
	\end{align*}
\end{subequations}
and
\begin{align*}
&\frac{\partial \mathcal{L}_{\rho} (\bm{z},\bm{\nu})}{\partial d_i} = \frac{e^{2 \gamma \left[ d_i + (t_0 - t_i)c \right]} - 1}{e^{2 \gamma \left[ d_i + (t_0 - t_i)c \right]} + 1} - \mu_{i+L+1} +\mu_{i+2L+1} + 2 \lambda_{i} d_i + \rho \Big\{ \mu_{i+L+1}^2 d_i\\
&~~+ \mu_{i+2L+1}^2 \left[ d_i - (t_i - t_0)c \right] + 2 \lambda_{i}^2 d_i \left( d_i^2 - {\| \bm{x} - \bm{x}_i \|}_2^2 \right) \Big\},~~i = 1,2,...,L.
\end{align*}

\section{Stability, convergence, and complexity analyses}
\label{CSCA}
Since the original objective function of our formulation lies in the $\ell_1$-space, we succinctly term the proposed projection-type recurrent neural network method $\ell_1$-PNN. In this section, several important aspects including the stability, convergence, and complexity properties of $\ell_1$-PNN are discussed.

\subsection{Local stability and convergence analysis}
As a preparation for the formal statements and by taking the GCOP (\ref{GCOP}) as an example, we define three concepts which frequently appear in the optimization literature:

\textbf{Definition 1.} (\texttt{Feasible region}). A feasible region is the set of all possible solutions to an optimization problem (namely, the GCOP (\ref{GCOP})) that satisfy the problem's constraints ($\bm{g}(\tilde{\bm{z}}) \leqq \bm{0}_K$ and $\bm{h}(\tilde{\bm{z}}) = \bm{0}_M$).

\textbf{Definition 2.} (\texttt{Regularity condition}). A feasible point $\tilde{\bm{z}}$ is said to be a regular point if the gradients of the active inequality constraints (i.e., $\bm{\nabla}_{\bm{z}} g_i (\tilde{\bm{z}}), \forall i \in \mathcal{I} = \left\{ i | g_i (\tilde{\bm{z}}) = 0 \right\}$) and those of the equality constraints (i.e., $\bm{\nabla}_{\bm{z}} h_i (\tilde{\bm{z}})$ for $i = 1,2,...,M$) are linearly independent at $\tilde{\bm{z}}$. This is a.k.a. the linear independence constraint qualification (LICQ).

\textbf{Definition 3.} (\texttt{Strict local minimum}). A point $\bm{z}^{*}$ is said to be a strict local minimum if $f(\bm{z}^{*}) < f(\bm{z}), \forall \bm{z} \in \mathcal{N}(\bm{z}^{*}, \delta) \cap \rm{S}$, where $\mathcal{N}(\bm{z}^{*}, \delta)$ represents the neighborhood of the point $\bm{z}^{*}$ with radius $\delta > 0$ and $\rm{S}$ denotes the feasible region.

A lemma presenting the second-order sufficient conditions (SOSC) \cite{MSBazaraa} is then introduced as:

\textbf{Lemma 1.} (\texttt{SOSC} \cite{MSBazaraa}). Let $\bm{z}^{*}$ be a feasible and regular point of the GCOP (\ref{GCOP}). If there exists a $\bm{\nu}^{*} = \left[ {\bm{\mu}^{*}}^T, {\bm{\lambda}^{*}}^T \right]^T \in \mathbb{R}^{K+M}$, such that $(\bm{z}^{*},\bm{\nu}^{*})$ is a KKT pair and $\bm{\nabla}_{\bm{z} \bm{z}}^{2} \bar{\mathcal{L}}(\bm{z}^{*},\bm{\nu}^{*})$ is positive definite on the cone
\begin{align*}
\mathcal{C} = \Big\{\bm{y} \in \mathbb{R}^N \Big| \left[ \bm{\nabla}_{\bm{z}} g_i (\bm{z}^{*}) \right]^T \bm{y} &= 0, \forall i \in \mathcal{I}_{+},~\left[ \bm{\nabla}_{\bm{z}} g_i (\bm{z}^{*}) \right]^T \bm{y} \leq 0, \forall i \in \mathcal{I}_{0},\\
\left[ \bm{\nabla}_{\bm{z}} h_i (\bm{z}^{*}) \right]^T \bm{y} &= 0, \forall i = 1,2,...,M,~\bm{y} \neq \bm{0}_N \Big\},
\end{align*}
where $\bar{\mathcal{L}}(\bm{z}^{*},\bm{\nu}^{*}) = f(\bm{z}^{*}) + \sum_{i \in \mathcal{I}} \mu_{i}^{*} g_i (\bm{z}^{*}) + \sum_{i = 1}^{M} \lambda_{i}^{*} h_i (\bm{z}^{*})$ is the restricted Lagrangian function at $(\bm{z}^{*},\bm{\nu}^{*})$, and $\mathcal{I}_{+} = \left\{ i \in \mathcal{I}|\mu_{i}^{*} > 0 \right\}$ and $\mathcal{I}_{0} = \left\{ i \in \mathcal{I}|\mu_{i}^{*} = 0 \right\}$ are often referred to as the sets of strongly active and weakly active constraints, respectively.

We now finally arrive at the following lemma in which the analytical results concerning the behaviors of iterative sequences produced by (\ref{dynamicsGCOP}) are established.

\textbf{Lemma 2.} (\texttt{Local stability} \cite{HChe}). Suppose that $(\bm{z}^{*},\bm{\nu}^{*})$ is a KKT point of the GCOP (\ref{GCOP}) satisfying the SOSC in Lemma 1. There exists a sufficiently large $\rho > 0$, such that the neurodynamic system described by (\ref{dynamicsGCOP}) is asymptotically stable at $(\bm{z}^{*},\bm{\nu}^{*})$, where $\bm{z}^{*}$ is a strict local minimum of the GCOP (\ref{GCOP}).

The detailed proof of Lemma 2 is omitted, because it constitutes a special case of the analysis of Theorem 2 in \cite{HChe} if we set the lower and upper bounds therein as negative and positive infinities, respectively. Based on Lemma 2, we embark on a careful examination of the local stability of $\ell_1$-PNN below. In general, the source onset time should be a proper value at least greater than $0$ \cite{ZSu}, the positions of the sensors are different from that of the source (otherwise there is no need for localization), and the positive bias error is much larger than the magnitude of the measurement noise in a TOA measurement under NLOS conditions \cite{GWang}. Therefore, the inequality constraints in (\ref{CLAD}) are actually all inactive (viz. $\mathcal{I} = \emptyset$), which means that the LICQ in our case is subject to only the equality constraints. The gradients of the equality constraints in (\ref{CLAD}) at a KKT point $(\bm{z}^{*},\bm{\nu}^{*})$ are calculated as
\begin{align}{\label{Gradh}}
\bm{\nabla}_{\bm{z}} \bm{h}(\bm{z}^{*}) &= \frac{\partial \bm{h}(\bm{z})}{\partial \bm{z}}\bigg|_{\bm{z} = \bm{z}^{*}} = \left[ \frac{\partial h_1 (\bm{z}^{*})}{\partial \bm{z}}, \frac{\partial h_2 (\bm{z}^{*})}{\partial \bm{z}},..., \frac{\partial h_{L} (\bm{z}^{*})}{\partial \bm{z}} \right]^T\nonumber\\
&= \begin{bmatrix}
\begin{array}{c : c : c}
\bm{0}_L  &  2 \left( \bm{X}^T - \bm{1}_L {\bm{x}^{*}}^T \right)  &  2\mathrm{diag}(\bm{d}^*)
\end{array}
\end{bmatrix}, 
\end{align}
where $\bm{1}_L \in \mathbb{R}^L$ is an all-one vector of length $L$, $\mathrm{diag}(\bm{a})$ stands for a diagonal matrix with vector $\bm{a}$ being its main diagonal, and $\bm{X} = \left[ \bm{x}_1, \bm{x}_2,..., \bm{x}_L \right] \in \mathbb{R}^{k \times L}$ represents a matrix including the positions of all sensors. Given the aforementioned practical considerations, we can easily deduce that the row vectors of the matrix in (\ref{Gradh}) are linearly independent and therewith $\mathcal{C} = \emptyset$. As a result, the SOSC hold trivially, and from Lemma 2 our $\ell_1$-PNN is assured locally stable as long as the Lagrangian parameter takes a large enough value. It is worth mentioning that due to the nonconvexity of the problem being solved, we investigate only the local stability of $\ell_1$-PNN here, but refer the interested readers to \cite{HChe,ZYan,HChe2} for the very recent developments of global convergence guaranteed neurodynamic optimization. Nevertheless, it is shown in Section \ref{SR} through extensive simulations that even local minimization can yield satisfactory performance in terms of positioning accuracy.

\subsection{Complexity analysis}

Since $\ell_1$-PNN is intended to be implemented analogously by designated hardware (e.g., application specific integrated circuits), it may not be meaningful to compare its complexity with those of the numerical approaches. Yet, we still manage to analyze the computational complexity of the neural network framework (\ref{dynamicsGCOP}) when it is realized in a discrete and numerical manner \cite{JLiang}:
\begin{equation}{\label{discrete}}
\left\{
\begin{aligned}
&\bm{z}_{(\kappa+1)} = \bm{z}_{(\kappa)} + \tau \frac{d\bm{z}}{dt},\\
&\bm{\mu}_{(\kappa+1)} = \bm{\mu}_{(\kappa)} + \tau \frac{d\bm{\mu}}{dt},\\
&\bm{\lambda}_{(\kappa+1)} = \bm{\lambda}_{(\kappa)} + \tau \frac{d\bm{\lambda}}{dt},
\end{aligned}
\right.
\end{equation}
where the subscript $(\cdot)_{(\kappa)}$ denotes the iteration index, $\tau$ is the step size, and the derivatives $\frac{d\bm{z}}{dt}, \frac{d\bm{\mu}}{dt}, \frac{d\bm{\lambda}}{dt}$ follow the definitions in (\ref{dynamicsGCOP}). With the help of Horner's scheme \cite{EHildebrand}, the evaluation of a polynomial of degree $n$ with fixed-size coefficients can be computed in $\mathcal{O}(n)$ time. Then, by considering polynomial evaluation as the operation in each step of (\ref{discrete}) governing the computational complexity, it is not hard to conclude that the dominant complexity of $\ell_1$-PNN is $\mathcal{O}\Big(N_{\text{PNN}}\Big(\max(\zeta,3)+5k+L\max(\zeta,5)+\frac{L^2 + 5L + 2}{2}+2L\Big)\Big) = \mathcal{O}\left(N_{\text{PNN}}L^2\right)$, where $\zeta$ is the degree of the Maclaurin polynomial for the hyperbolic tangent function $\frac{e^{2 \gamma \left[ d_i + (t_0 - t_i)c \right]} - 1}{e^{2 \gamma \left[ d_i + (t_0 - t_i)c \right]} + 1}$ and $N_{\text{PNN}}$ is the iteration number of the PNN using discrete realization. Table \ref{table} presents a comparison of complexity of the proposed neurodynamic method for solving (\ref{CLAD}) (termed $\ell_1$-PNN), SDP-based robust method for solving Formulation 1 in \cite{GWang3} (termed SDP-Robust-Refinement-1), SDP-based robust method for solving Formulation 2 in \cite{GWang3} (termed SDP-Robust-Refinement-2), and SDP-based model transformation method in \cite{ZSu} (termed SDP-TOA) as the function of $L$. Note that the naming of SDP-TOA is consistent with that in \cite{GWang3}, and the computational costs of dealing with the mixed SDP/SOCP problems are determined by following the calculation rule in \cite{ABenTal}. It can be concluded that $\ell_1$-PNN has a significantly lower complexity than those convex optimization approaches in \cite{GWang3,ZSu}.

\begin{table*}[!t]
	\renewcommand{\arraystretch}{0.6}
	\caption{Complexity of considered NLOS mitigation algorithms}
	\label{table}
	\centering
	\begin{tabular}{c|c}
		\hline\hline
		\bfseries Algorithm & \bfseries Complexity\\
		\hline
		$\ell_1$-PNN & $\mathcal{O}\left(N_{\text{PNN}}L^2\right)$\\
		\hline
		SDP-Robust-Refinement-1 & $\mathcal{O}\left(L^{6.5}\right)$\\
		\hline
		SDP-Robust-Refinement-2 & $\mathcal{O}\left(L^{6.5}\right)$\\
		\hline
		SDP-TOA & $\mathcal{O}\left(L^{4}\right)$\\
		\hline\hline
	\end{tabular}
\end{table*}

\section{Simulation results}
\label{SR}
This section substantiates the efficacy of our proposed neurodynamic approach through simulation studies. To be specific, $\ell_1$-PNN is compared with representative NLOS mitigation algorithms including SDP-Robust-Refinement-1 in \cite{GWang3}, SDP-Robust-Refinement-2 in \cite{GWang3}, and SDP-TOA in \cite{ZSu} just as what have been provided in Table \ref{table}\footnote{It is remarkable that the definition of matrix $\bm{E}$ in \cite{GWang3} is incorrect and should be amended before putting the involved algorithms into use.}, and additionally the separated constrained weighted LS (SCWLS) approach in \cite{LLin}. Furthermore, the Cram\'{e}r-Rao lower bounds (CRLBs) for positioning with TDOA measurements in the LOS \cite{LLin} and NLOS \cite{YQi} scenarios are also included as the benchmark (when applicable). It should be mentioned that the invocation of $\ell_1$-PNN and SDP-TOA needs only the sensor positions and known signal timestamps as the inputs, whereas additional prior knowledge of the error bound/noise variance is a must for SDP-Robust-Refinement-1, SDP-Robust-Refinement-2, and SCWLS. In the following numerical examples, a perfect upper bound of the NLOS error is always ensured and passed into SDP-Robust-Refinement-1 and SDP-Robust-Refinement-2. The CVX package \cite{MGrant} and MATLAB$^\circledR$ ODE solver are utilized for realizing the convex programs and solving the systems of equations, respectively. All hyperparameters involved in SDP-TOA are assigned the same values as those in the demonstration program\footnote{\url{https://github.com/xmuszq/Semidefinite-Programming-SDP-optimization}} coded by the authors of \cite{ZSu}. As a global setup of $\ell_1$-PNN, the values held in the variable and Lagrangian neurons are initialized with 0s. For the selection of the augmented Lagrangian parameter $\rho$, the existing numerical results \cite{HChe} demonstrate that a relatively large $\rho$ can reduce transient oscillation of the neurodynamic model and speed up the convergence. In our simulations, we simply set $\rho = 5$ and it is observed that such a value always makes $\ell_1$-PNN settle down within several tens of time constants. Another predefined parameter associated with the quality of approximation to the original $\ell_1$-norm is fixed as $\gamma = 100$, based on which the resultant estimator is robust enough (see Fig. \ref{xiong3}). All simulations are carried out using a laptop with Intel$^\circledR$ Core$^\text{TM}$ i7-10710U processor and 16 GB memory.

Basically, two representative configurations with $k = 2$ are covered. The first configuration considers source localization in a 20 m $\times$ 20 m square region with $L=8$ sensors being evenly placed on the perimeter of the area and a single source being deployed at $\bm{x}=[2, 3]^T$ m. On the other hand, a typical setting in \cite{WXiong} with multiple sensors and a single source, whose locations are all randomly selected from the 20 m $\times$ 20 m square region in each Monte Carlo (MC) run, is adopted as the second configuration. The true value of the unknown source onset time is fixed as $t_0 = 0.1$ s, while the known signal timestamps received at the sensors and the TDOA measurements in the simulated system are obtained in accordance with (\ref{TOA}) and (\ref{TDOA}), respectively. Particularly, the signal propagation speed is set as $c=1$ m/s to keep things simple, the zero-mean Gaussian distributed noise $n_i$ is assumed to be of identical variance $\sigma^2$ for all $i$s, and the possible NLOS error in the TOA measurement between the source and $i$th sensor, namely $q_i$, is generated from the uniform distribution\footnote{Unquestionably, the corresponding source-sensor path is LOS if $\omega_i$ is assigned $0$.} $\mathcal{U}(0,\omega_i)$.

\begin{figure*}[!t]
	\centering
	\subfigure[]{\includegraphics[width=2.36in]{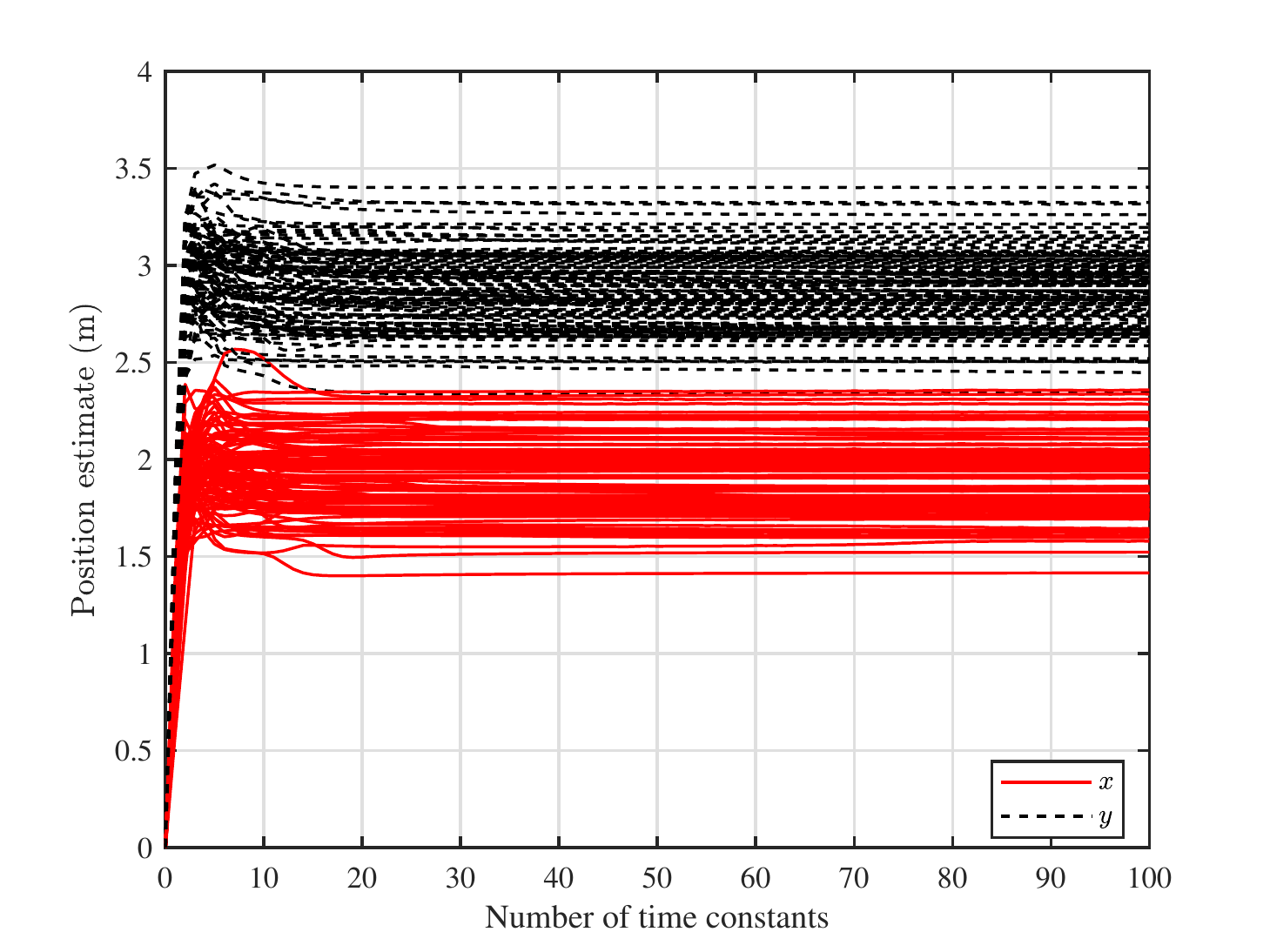}}%
	\subfigure[]{\includegraphics[width=2.36in]{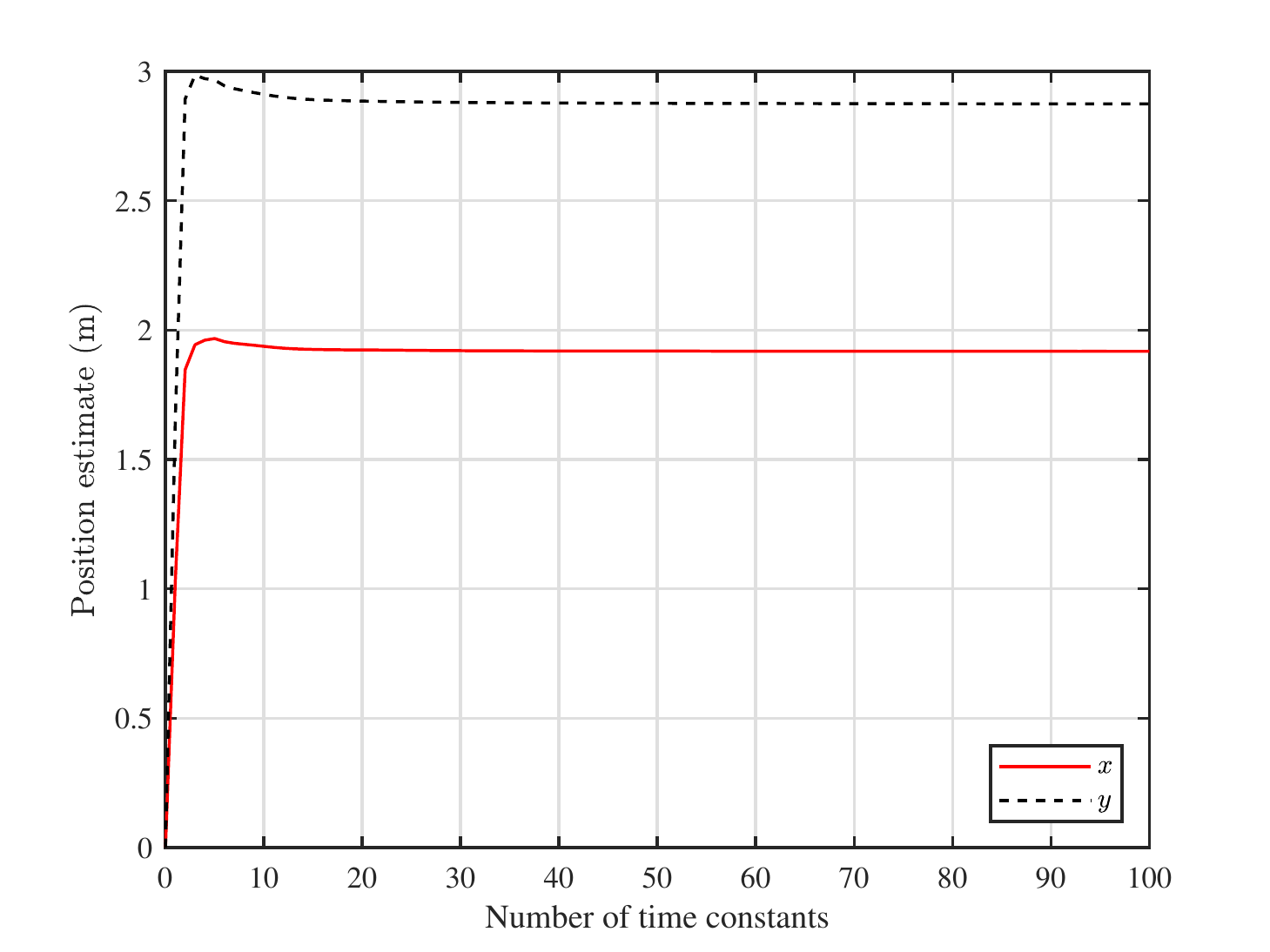}}
	\subfigure[]{\includegraphics[width=2.36in]{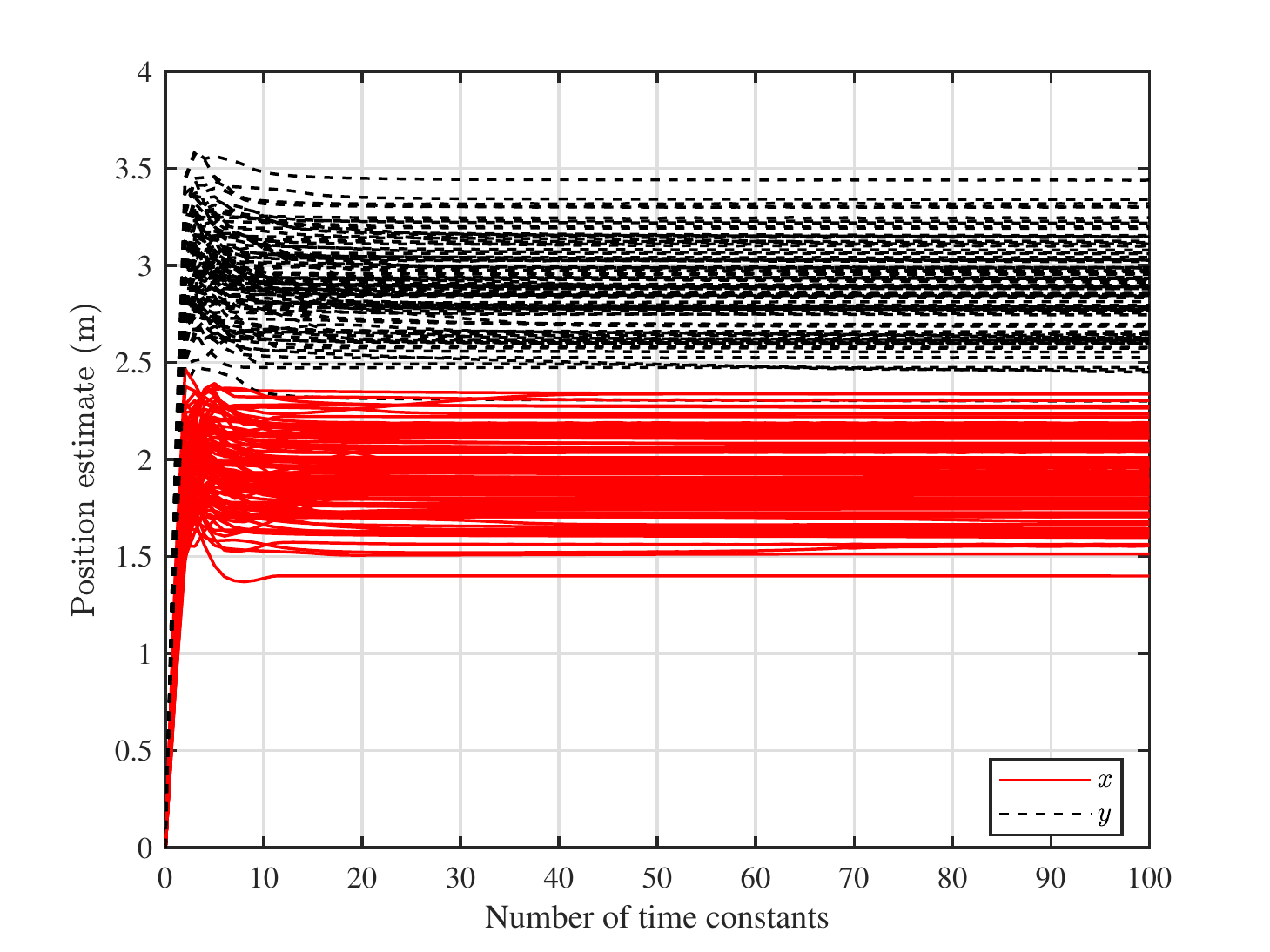}}
	\subfigure[]{\includegraphics[width=2.36in]{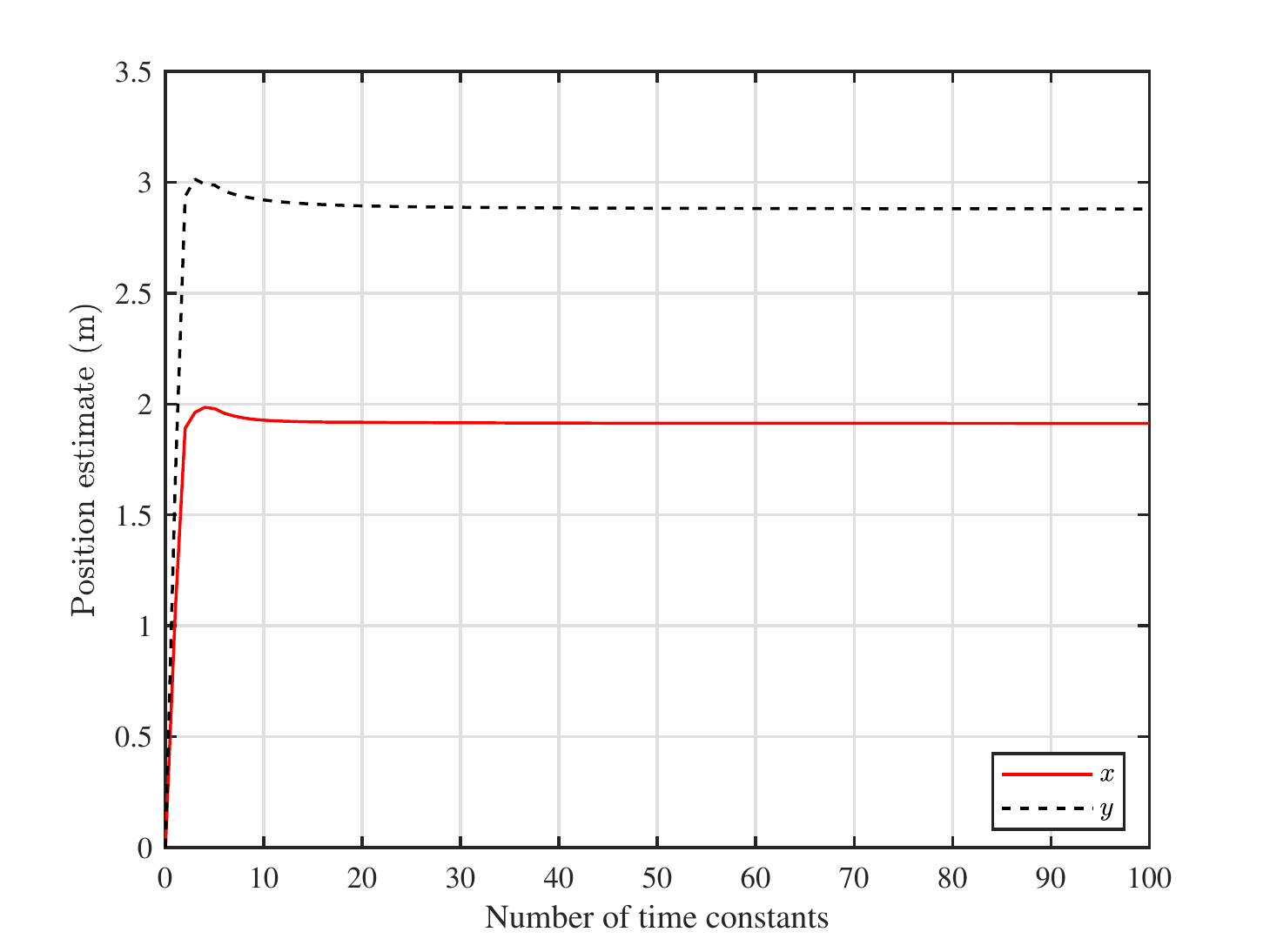}}
	\centering
	\caption{Dynamic behaviors of estimated source position versus time constant number for deterministic deployment in LOS and mild NLOS environments. (a) Outputs of 100 independent trials when $\sigma^2 = 0.1$ and $\omega_i = 0$ for all $i$s. (b) Mean of 100 outputs when $\sigma^2 = 0.1$ and $\omega_i = 0$ for all $i$s. (c) Outputs of 100 independent trials when $\sigma^2 = 0.1$, $\omega_1 = 5$, $\omega_5 = 5$, and $\omega_i = 0$ for other $i$s. (d) Mean of 100 outputs when $\sigma^2 = 0.1$, $\omega_1 = 5$, $\omega_5 = 5$, and $\omega_i = 0$ for other $i$s.}
	\label{xiong5678}
\end{figure*}

\begin{figure*}[!t]
	\centering
	\includegraphics[width=3.5in]{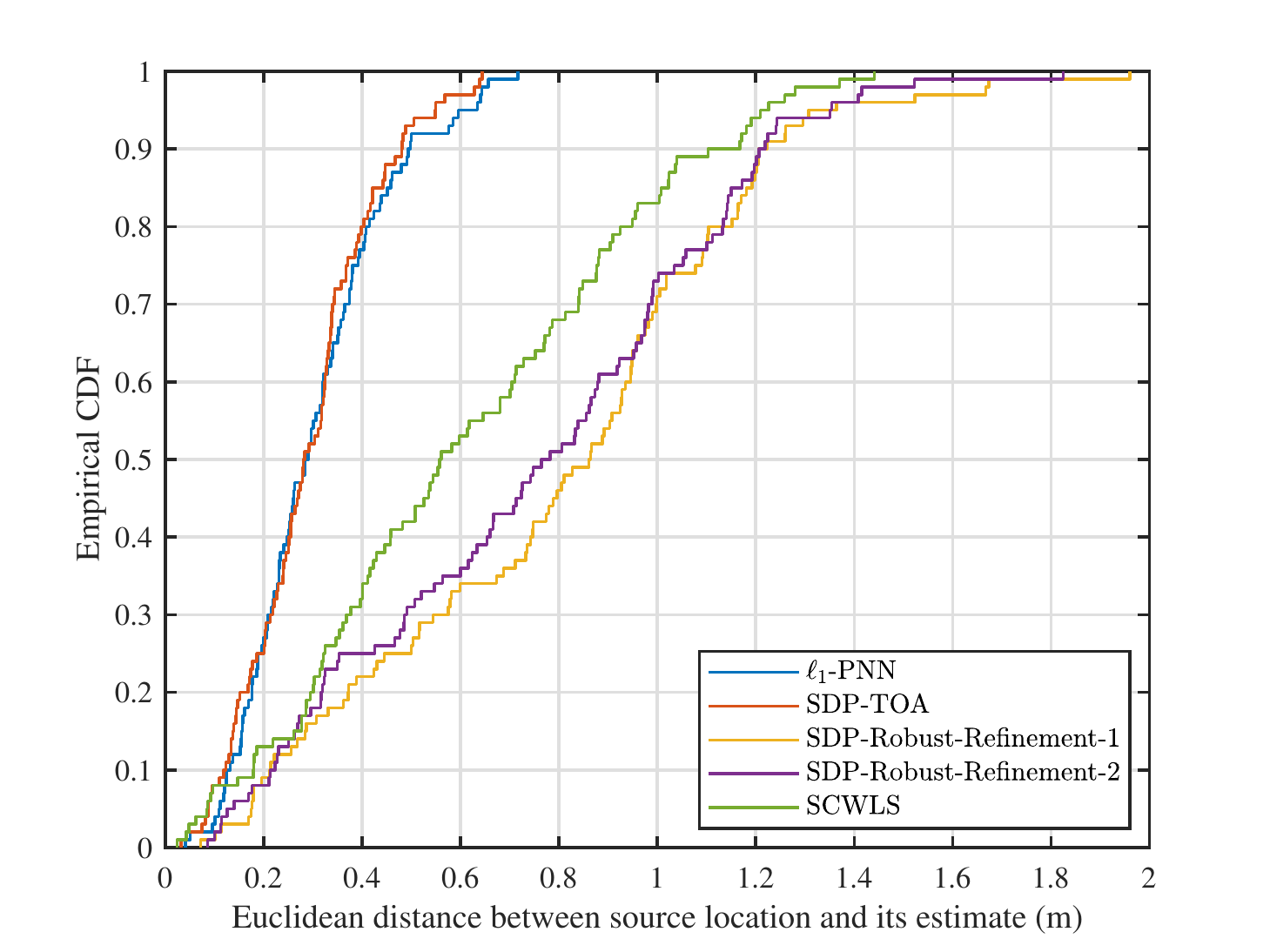}
	\centering
	\caption{Empirical CDF of Euclidean distance between source location and its estimate for deterministic deployment in mild NLOS environment based on 100 MC runs when $\sigma^2 = 0.1$, $\omega_1 = 5$, $\omega_5 = 5$, and $\omega_i = 0$ for other $i$s.}
	\label{ECDF}
\end{figure*}

\begin{figure*}[!t]
	\centering
	\includegraphics[width=3.5in]{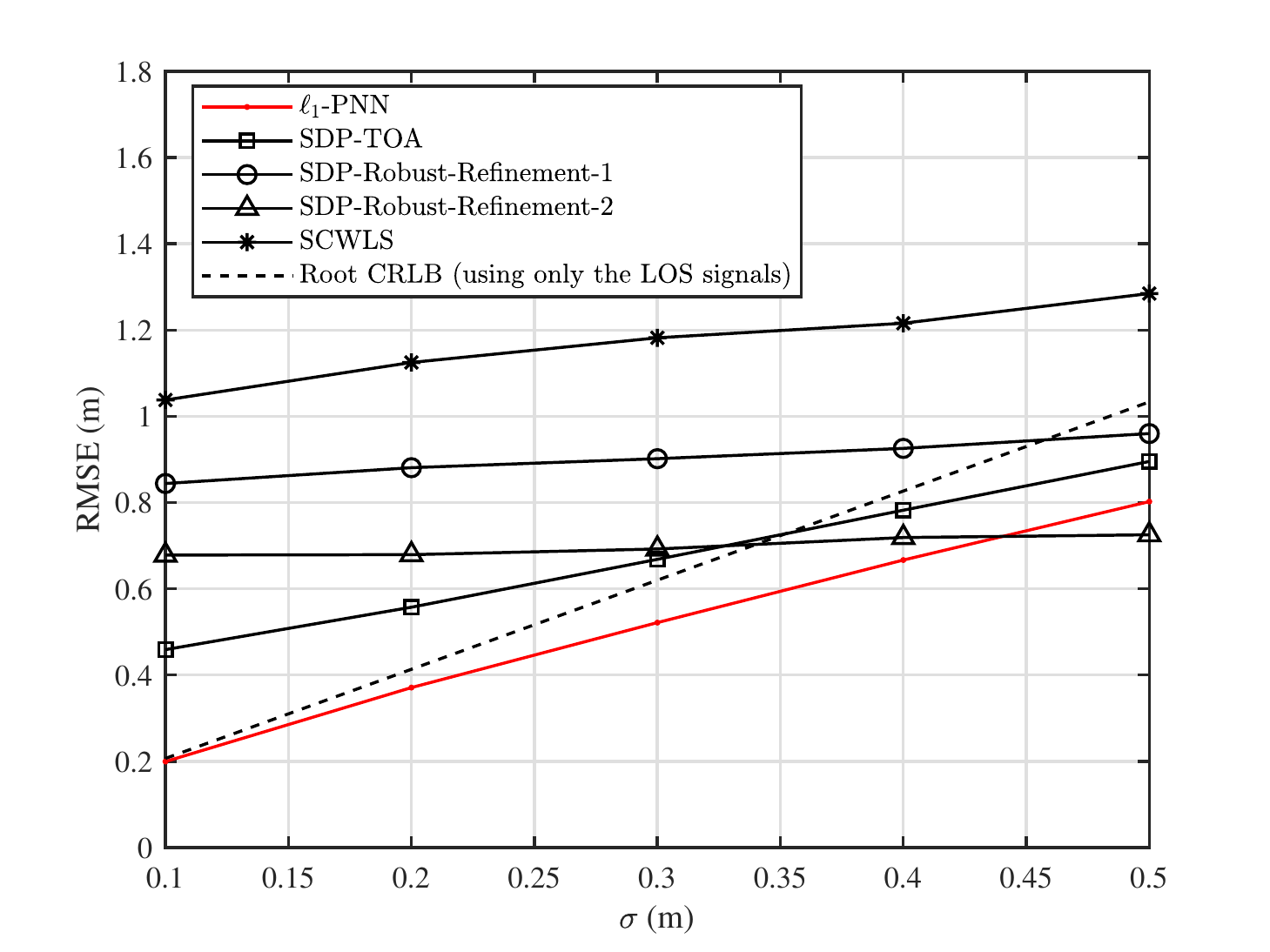}
	\centering
	\caption{RMSE versus $\sigma$ for deterministic deployment in mild NLOS scenario when $\omega_1 = 5$, $\omega_2 = 5$ and $\omega_i = 0$ for other $i$s.}
	\label{NLOSCRLB}
\end{figure*}

\begin{figure*}[!t]
\centering
\includegraphics[width=3.5in]{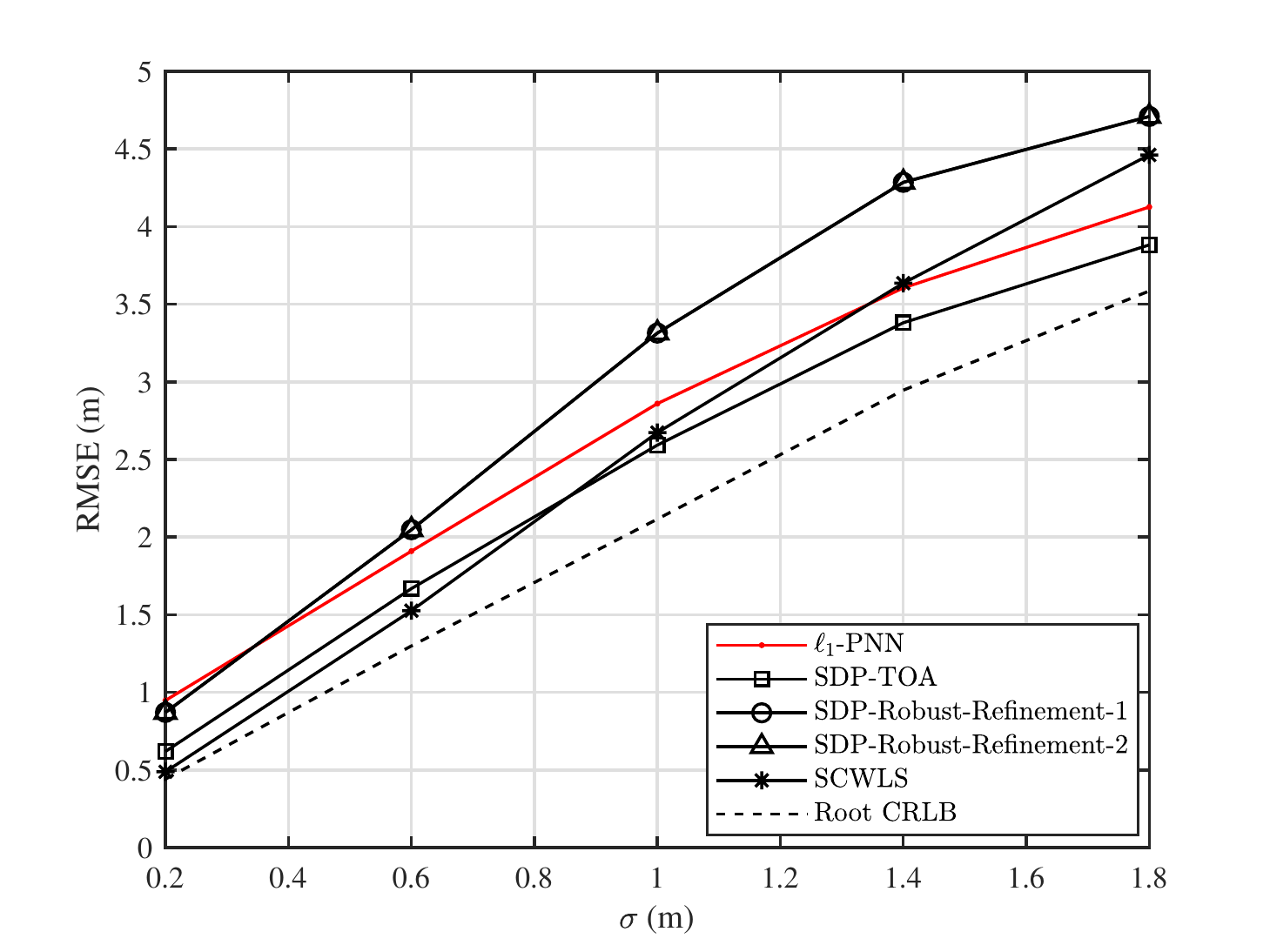}
\caption{RMSE versus $\sigma$ in LOS scenario (viz. $L_{\textup{NLOS}} = 0$).}
\label{xiong9}
\end{figure*}

\begin{figure*}[!t]
	\centering
	\subfigure[]{\includegraphics[width=2.36in]{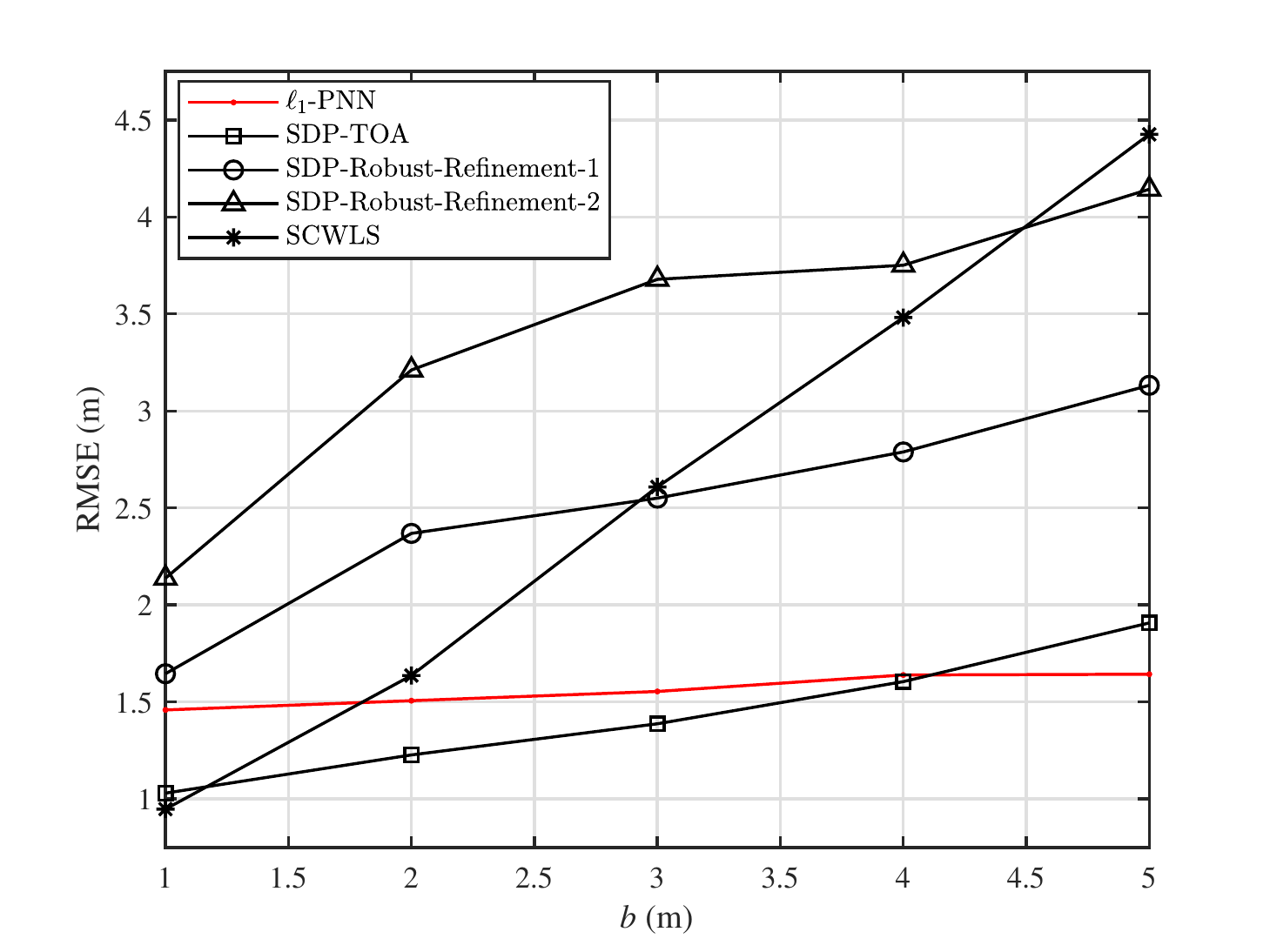}}
	\subfigure[]{\includegraphics[width=2.36in]{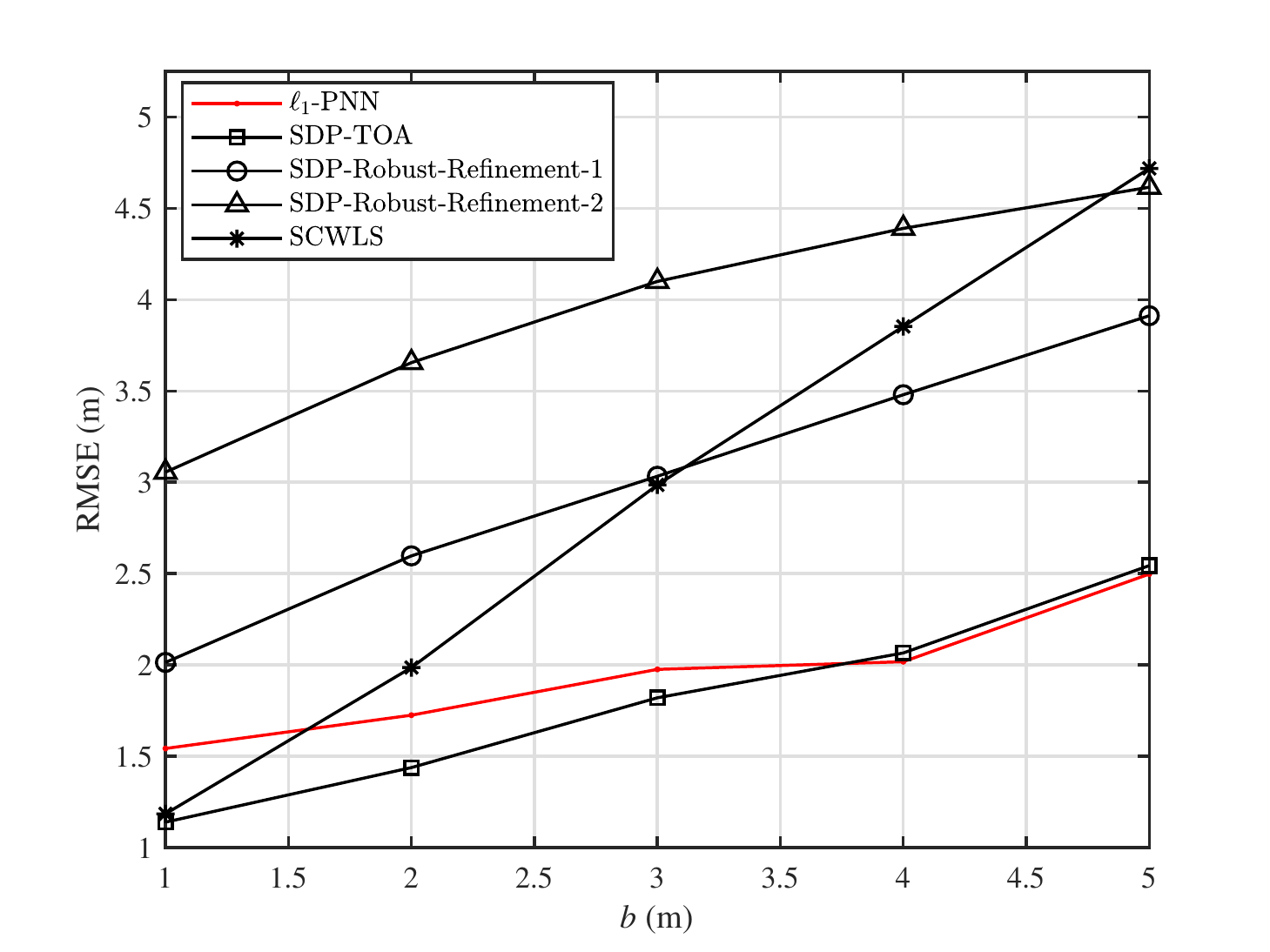}}
	\subfigure[]{\includegraphics[width=2.36in]{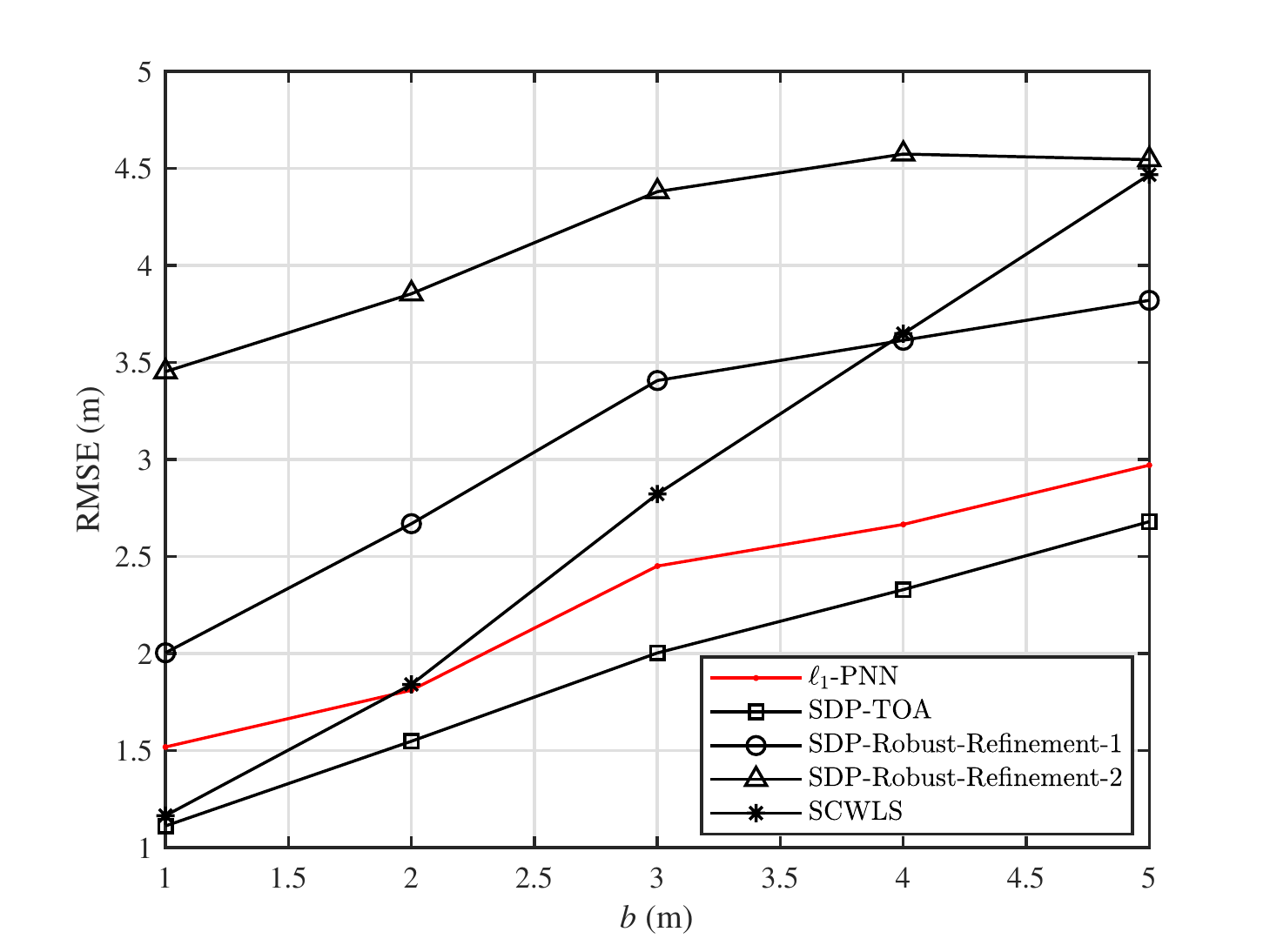}}
	\subfigure[]{\includegraphics[width=2.36in]{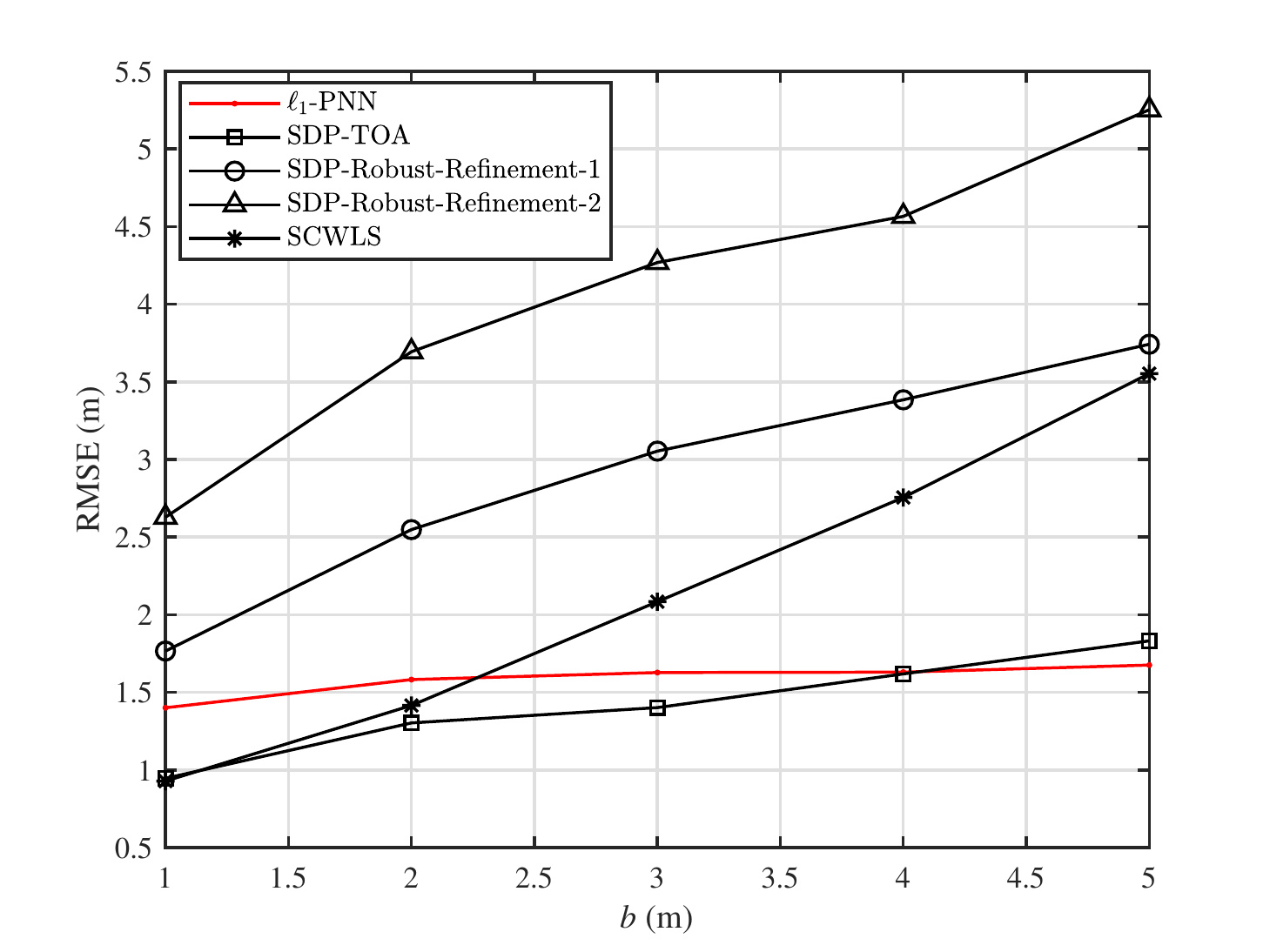}}
	\subfigure[]{\includegraphics[width=2.36in]{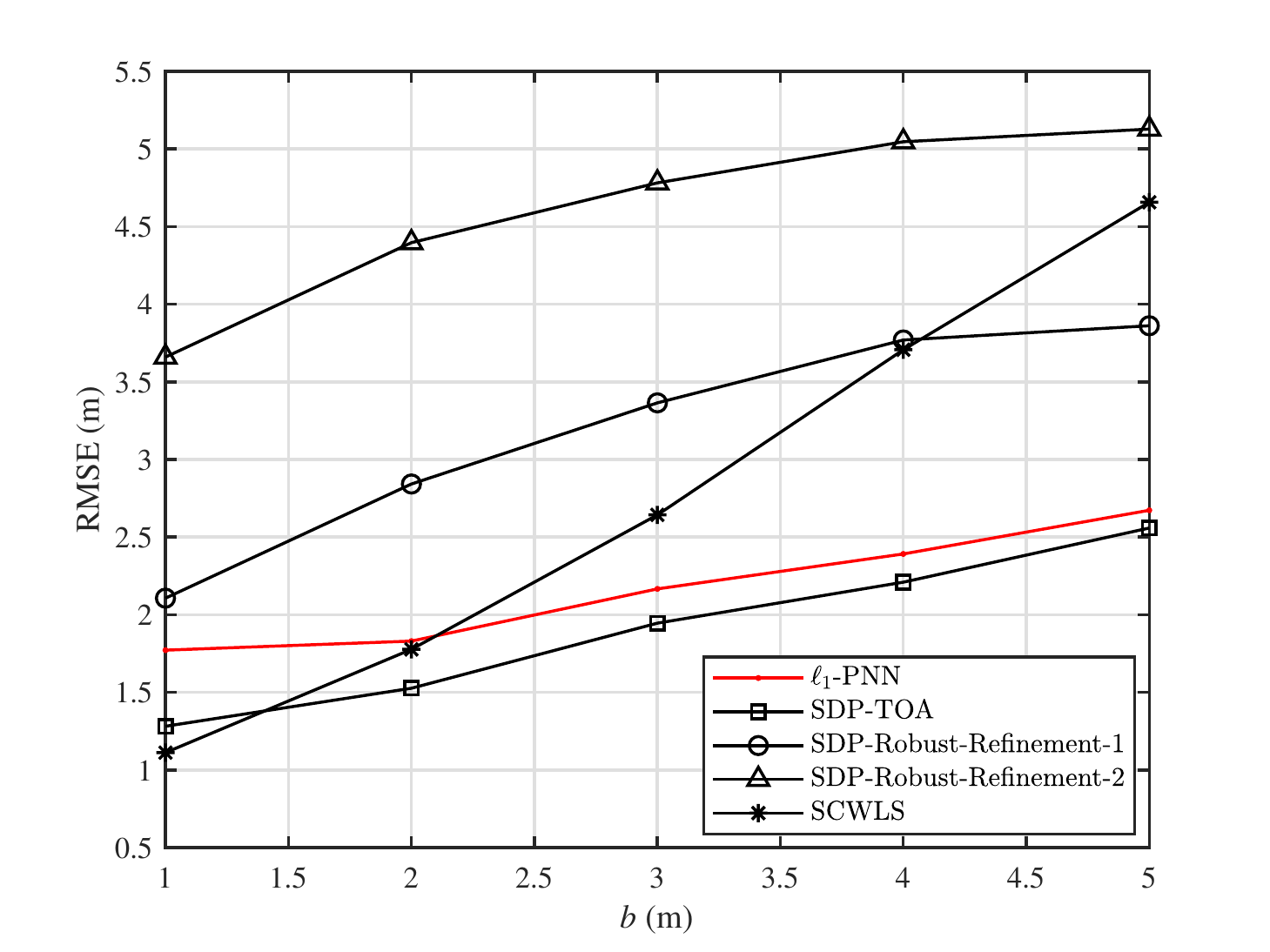}}
	\subfigure[]{\includegraphics[width=2.36in]{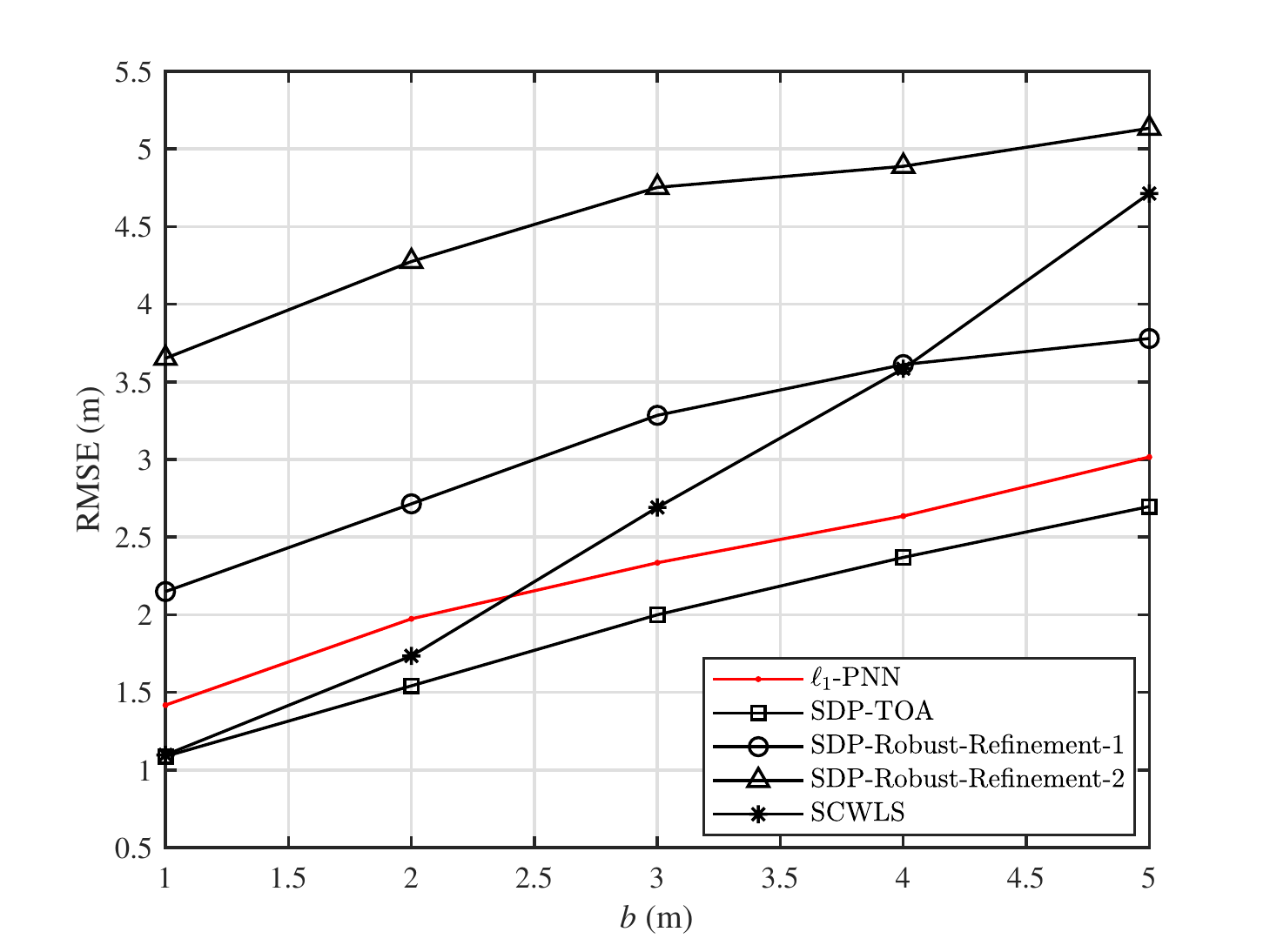}}
	\centering
	\caption{RMSE versus parameter of uniform distribution $b$ in different NLOS scenarios when $\sigma^2 = 0.1$. (a) $L_{\textup{NLOS}} = 2$, $\omega_1 = b$. (b) $L_{\textup{NLOS}} = 5$, $\omega_1 = b$. (c) $L_{\textup{NLOS}} = 8$, $\omega_1 = b$. (d) $L_{\textup{NLOS}} = 2$, $\omega_1 = 0$. (e) $L_{\textup{NLOS}} = 5$, $\omega_1 = 0$. (f) $L_{\textup{NLOS}} = 8$, $\omega_1 = 0$.}
	\label{xiong10to15}
\end{figure*}

In the first test, the dynamic behaviors of the estimated source position using $\ell_1$-PNN in the deterministic deployment scenario are investigated. Taking the LOS and a mild NLOS environments for instance, Fig. \ref{xiong5678} plots the dynamics of the second and third variable neurons (i.e., those holding the variable $\bm{x}$) based on 100 MC runs. It is seen that $\ell_1$-PNN settles down and converges to a point close to the true source location within 20 to 40 time constants. For this reason, in the following we simply take the corresponding neuron output right after 40 time constants as the final position estimate produced by $\ell_1$-PNN. As a preliminary evaluation of $\ell_1$-PNN in comparison with other considered methods, Fig. \ref{ECDF} shows the empirical cumulative distribution function (CDF) of the Euclidean distance between source location and its estimate in the above-defined mild NLOS environment, from which we see that $\ell_1$-PNN and SDP-TOA demonstrate superior positioning performance. Next, the root mean square error (RMSE) criterion with 500 ensemble trials, defined as $\textup{RMSE} = \sqrt{\frac{1}{500}\sum_{i=1}^{500}{{\left\|\hat{\bm{x}}^{\{i\}} - \bm{x}^{\{i\}}\right\|}^2}}$ where $\hat{\bm{x}}^{\{i\}}$ represents the estimate of source position in the $i$th MC run (namely ${\bm{x}}^{\{i\}}$), is utilized as a measure to further compare the location estimation performance of diverse approaches. Fig. \ref{NLOSCRLB} depicts the RMSE versus $\sigma$ for the deterministic deployment scenario when the number of NLOS connections is fixed as $L_{\textup{NLOS}} = 2$ and the parameter of uniform distribution is set to 5. Especially, comparison with the CRLB when no prior NLOS statistics are available (namely, the one depending only on LOS signals \cite{YQi}) is also made. The results reveal that: (i) $\ell_1$-PNN exhibits the best robustness to NLOS propagation in such circumstances as long as $\sigma$ is not large enough, and (ii) taking advantage of rather than simply discarding the NLOS links results in an improvement in performance.

The random deployment scenario with $L=10$ is now considered to assess the localization performance of $\ell_1$-PNN together with other state-of-the-art algorithms under LOS and NLOS conditions. It must be pointed out that the setup is quite different from and in one sense more general than those in \cite{GWang2} and \cite{GWang3}, as the sensors here are neither fixed nor placed on a certain circle but all randomly drawn from the square region. Fig. \ref{xiong9} illustrates the RMSE as a function of $\sigma$ in the scenario where LOS transmissions are guaranteed for all source-sensor paths, i.e., $L_{\textup{NLOS}} = 0$. Clearly, only the SCWLS algorithm at sufficiently lower-level measurement disturbances (e.g., when $\sigma = 0.2$ m) can attain the CRLB \cite{LLin}. On the other side, there is always a performance gap between $\ell_1$-PNN and SDP-TOA/CRLB (i.e., SDP-TOA and CRLB are superior to $\ell_1$-PNN by about $0.25$ m and $0.5$ m across the whole range of $\sigma$). This is owing to the fact that SDP-TOA tightly approximates the ML estimator for small noise of the same level, whereas $\ell_1$-PNN derived in $\ell_1$-space is inherently suboptimal under the Gaussian noise assumption. It is also observed that the two worst-case robust methods SDP-Robust-Refinement-1 and SDP-Robust-Refinement-2 in general perform badly in the LOS scenario. We divide the test conditions in scenarios where NLOS propagation exists into two separate groups: (i) the path between the source and reference sensor is NLOS, and (ii) the path between the source and reference sensor is LOS. In each group, three diverse cases with $L_{\textup{NLOS}}=2,5,8$ are included, standing for the mild, moderate, and severe NLOS environments, respectively. Fig. \ref{xiong10to15} shows the comparison results with the detailed parameter settings being given in the caption. We can see that the location estimation accuracy of the non-robust SCWLS scheme deteriorates considerably as $b$ increases. $\ell_1$-PNN and SDP-TOA have comparable RMSEs, and they both outperform SDP-Robust-Refinement-1 and SDP-Robust-Refinement-2. Note that although $\ell_1$-PNN is slightly inferior to SDP-TOA in most cases (e.g., for $b < 4$ in Figs. \ref{xiong10to15}(a), \ref{xiong10to15}(b), and \ref{xiong10to15}(d) and all $b$s in Figs. \ref{xiong10to15}(c), \ref{xiong10to15}(e), and \ref{xiong10to15}(f)), the former is computationally more efficient and gets rid of the cumbersome hyperparameter tuning problems.

\section{Conclusion}
\label{CC}
In this paper, we proposed a robust model transformation formulation for TDOA-based source localization and devised a novel neurodynamic optimization solution to it. The new scheme does not require any \textit{a priori} information except the positions of the sensors, received signal timestamps thereat, and signal propagation speed, as the mitigation of NLOS biases in the reconstructed TOA measurements are achieved via the $\ell_1$-norm criterion. To address the problem of non-differentiability of the $\ell_1$-norm, certain approximations were applied to the original objective function for yielding a differentiable surrogate. Benefiting from the use of a projection-type recurrent neural network approach, the biggest advantage of the presented algorithm over the existing ones is its quadratic computational complexity in $L$. Through the theoretical analysis and extensive simulation investigations, we verified that the dynamics of the proposed $\ell_1$-norm-based PNN are locally stable, and confirmed its superiority over several existing TDOA-based localization schemes in terms of the estimation accuracy.

\section*{Acknowledgment}
This work was supported by the German Federal Ministry of Education and Research (BMBF) in the framework of the ASSIST~ALL project under Grant FKZ:16SV8162.

The first author would also like to thank Ms Ge Cheng at China Railway Major Bridge Reconnaissance \& Design Institute Co., Ltd. (BRDI), the team of Telocate GmbH, and the Laboratory for Electrical Instrumentation of the University of Freiburg for their assistance with the manuscript preparation.



\end{document}